\documentclass[showpacs,aps, prb,preprint,floatfix,superscriptaddress]{revtex4}
\begin{document}
\title{
Theory of the spin singlet filling factor $\nu=2$  quantum Hall droplet}
\author{Andreas Wensauer}
\affiliation{Institute for Microstructural Sciences, National Research
         Council Canada, Ottawa, K1A 0R6}
\affiliation{Institut f\"ur Theoretische Physik, 
         Universit\"at Regensburg, D-93040 Regensburg, Germany}
\author{Marek Korkusinski}
\affiliation{Institute for Microstructural Sciences, National Research
         Council Canada, Ottawa, K1A 0R6}
\author{Pawel Hawrylak}
\affiliation{Institute for Microstructural Sciences, National Research
         Council Canada, Ottawa, K1A 0R6}
\date{\today}


\begin{abstract}
A theory of electronic properties of a spin-singlet quantum Hall droplet at 
filling factor $\nu=2$ in a parabolic quantum dot is developed. 
The excitation spectrum and the stability of the droplet due to the transfer 
of electrons into the second Landau level at low magnetic fields and due to
spin flip at the edge at higher magnetic fields is determined using
Hartree-Fock, exact diagonalisation, and spin-density functional methods.
We show that above a critical number of electrons $N_c$ the
unpolarised $\nu=2$ quantum Hall droplet ceases to be a 
ground state in favor of spin-polarised phases.
We determine the characteristic pattern in the addition and current-amplitude
Coulomb blockade spectra associated with the stable $\nu=2$ droplet.
We show that the spin transition of the droplet at a critical number of 
electrons is accompanied by the reversal of the current amplitude modulation 
at the $\nu=2$ line, as observed in recent experiments.
\end{abstract}

\pacs{73.21.-b, 85.35.-p}

\maketitle

\section{Introduction}
In strong magnetic fields, electrons confined in quasi-two-dimensional
quantum dots form quantum Hall droplets (QHD).\cite{dot reviews}
The simplest examples of QHDs are the spin polarised $\nu=1$ droplet and the
spin singlet $\nu=2$ droplet.
The $\nu=2$ QHD corresponds to a droplet of electrons occupying the increasing
in energy spin-up and spin-down states of the lowest Landau level 
(LLL).\cite{ciorga00} 
QHDs have been extensively investigated experimentally\cite{ciorga00,tarucha00,%
hawrylak99,austing99,tarucha96,ray secs,mceuen,klein,osterkamp,sachrajda,marcus,weis,blick} and
theoretically.\cite{ciorga00,hawrylak99,mceuen,pawel secs,allan dot,eric allan mike,%
wen,oaknin,imamura,pawel dot imp,arek spectral}
Theoretically, the QHD at filling factor $\nu=1$ has attracted most 
attention.\cite{allan dot,eric allan mike,wen,klein}
Experimentally, however, the $\nu=1$ spin polarised droplet is not easily 
identified in the Coulomb blockade (CB) addition spectrum.\cite{osterkamp}
By contrast, the most pronounced feature in the addition spectrum of quantum 
dots, the $\nu=2$ line, is believed to originate from the formation of the 
$\nu=2$ QHD.\cite{ray secs}
Not surprisingly, recent experiments in quantum dots with controlled electron 
numbers,\cite{tarucha00} and controlled electron numbers combined with spin 
polarised injection/detection\cite{ciorga00} concentrated on the $\nu=2$ QHD.
Such a droplet is an  example of a chiral Fermi liquid, with both charge and
spin excitations, and as such can shed light on the current problem of spin 
and charge separation in correlated electron systems. 
Because this droplet is unstable at low magnetic fields against the transfer 
of electrons to the higher Landau level, it offers new spectroscopic 
opportunities as well as a more stringent test of various approximations to 
the quantum dot problem.
Despite experimental interest and theoretical opportunities, the detailed 
physical understanding of the $\nu=2$ QHD in terms of electron numbers, 
confining energy, magnetic field, and characteristic CB spectrum is rather 
limited.
In this paper we fill this gap and develop a theory of spin singlet QHD  
at $\nu=2$ in parabolic quantum dots. 
The parabolic confinement is chosen because it is a general feature of soft 
confining potentials in gated lateral and vertical devices, as well as in 
lens shaped self-assembled quantum dots.
The excitation spectrum and the stability of the droplet as a function
of the magnetic field, electron number $N_e$, confining energy $\omega_0$, and
Zeeman energy $E_z$ is determined. 
The spin-singlet $\nu=2$ droplet is found to be unstable due to the
transfer of electrons into the second Landau level at low magnetic fields 
and due to spin flip at the edge at higher magnetic field.  
Using Hartree-Fock, spin-density functional, and exact diagonalisation methods
we determine the stability conditions of the droplet, and the characteristic 
addition and current amplitude CB spectrum. 
We show that above a critical number of electrons $N_c$, the unpolarised 
$\nu=2$ quantum-Hall droplet ceases to be a ground state in favor of 
spin-polarised phases. 
The signature of this  transition turns out to be almost invisible in the 
position of CB peaks, but results in the reversal of the
current-amplitude modulation at the $\nu=2$ line with increasing
particle number, as observed in recent experiments.\cite{ciorga01} 
We find this reversal to be very sensitive to electronic correlations.

Our paper is organized as follows: 
In Sec.\ II we  define the single particle spectrum, the interacting
system, and the $\nu=2$ QHD. 
In Sec.\ III we analyse the properties of the $\nu=2$ QHD in the lowest Landau
level approximation. 
We calculate the charge and spin excitation spectrum, derive the expressions
describing conditions for electron transfer to the second Landau level and for 
the spin flip at the edge, and calculate the phase diagram. 
We end with the  calculation of the pattern of the addition spectrum 
characteristic to the $\nu=2$ QHD.
In Sec.\ IV we extend the calculation of the phase diagram 
to include Landau level mixing within Hartree-Fock, exact 
diagonalisation using one pair excitations, and spin-density functional 
theory in local density approximation (LSDA).
Sec.\ V is devoted to amplitude modulation patterns expected in spin
blockade (SB) spectroscopy. 
We identify sets of configurations responsible for SB amplitude modulation 
patterns both in the $\nu=2$ phase and in the region of the phase diagram 
corresponding to the breakdown of the spin singlet droplet. 
We show that only inclusion of correlations allows us to obtain reversal of 
amplitude modulation along the $\nu=2$ line observed in experiment.
Sec.\ VI summarizes our results.

\section{The model}

\subsection{Single particle states and non-interacting electron picture}

The  energy spectrum $E_{mn\sigma}$ and eigenstates $|m,n,\sigma\rangle$ 
of an electron localized on a quasi-two-dimensional parabolic quantum
dot are that of two harmonic oscillators in states $m,n$:   
$E_{mn\sigma}= \Omega_+\left(n+\frac{1}{2}\right) 
+ \Omega_-\left(m+\frac{1}{2}\right) + g \mu B \sigma$, and 
$|m,n;\sigma\rangle=\sqrt{\frac{1}{m!n!}} (a^+)^m (b^+)^n |0,0;\sigma\rangle$,
where  $a^+, b^+$ are   harmonic oscillator 
operators and  $\sigma$ denotes electron  spin \cite{pawel secs}.
The two harmonic frequencies are $\Omega_{\pm}= (\Omega \pm\omega_c)/2$
($\hbar=1$ for the rest of this work), $\omega_c=e B/ m^* c $ is the 
cyclotron energy, $l_{0}=1/(m^*\omega_c)^{1/2}$ is the magnetic length, 
$m^*$ is the effective mass, $\Omega= \sqrt{\omega_c^2+4\omega_0^2}$, 
and $g \mu B =E_z$ is the Zeeman energy.
With increasing magnetic field the energy $\Omega_{-}$ decreases to zero  
while $\Omega_{+}$ increases and approaches cyclotron energy.
When the magnetic field increases, the states $|m,n=0\rangle$ evolve into
the states of the lowest Landau level while the states $|m,n=1\rangle$ evolve
into the states of the second Landau level.
The single particle energy levels $E(m,0,\sigma)$ and $E(0,1,\sigma)$ as a 
function of the magnetic field B are shown in Fig.~\ref{dot0}. 
For illustration purposes a very high Zeeman energy of $E_z=0.15$ meV/T, 
comparable to the kinetic energy quantization $\Omega_{-}$, was used. 
We see that in high magnetic fields there are low energy states 
$|m,0,\pm\rangle$ lower than the lowest energy orbital $|m=0,n=1\rangle$ 
of the second Landau level.
They form a ladder of states with energies 
$E_{m,0,\pm}= \Omega_+\left(0+\frac{1}{2}\right) 
+ \Omega_-\left(m+\frac{1}{2}\right) \pm \frac{1}{2} g \mu B $.
This ladder of states is  marked by bars in Fig.\ref{dot0}. 
When these lowest energy states are occupied by $N_e= 2 N$  electrons, a 
finite $\nu=2$ spin-singlet chiral Fermi liquid droplet is formed as shown
in Fig.~\ref{dot0}b. 
In additon to occupied and empty lowest Landau level orbitals, there is a 
parallel ladder of second Landau level states $|m,n=1\rangle$ at higher
energy. 
Decreasing the magnetic field lowers the energy of the $|m=0,n=1\rangle$
state with respect to the energy of the highest occupied $|m=N-1,0\rangle$ 
orbital of the lowest Landau level, and an electron transfer occurs. 
The $|0,1\rangle$ state to which an electron transfers corresponds to
an orbital localised in the center of the dot, while the highest occupied state 
$|m,0\rangle$  corresponds to an orbital localised at the edge of the droplet. 
This is illustrated in Fig.~\ref{wf}(a) which shows charge distributions 
for the states $|0,1\rangle$ and $|9,0\rangle$.
In Fig.~\ref{wf}(b) we show  the total charge distribution of 14 electrons in 
the $\nu=2$ droplet, the distribution after an electron transfers to the center,
and the difference of charge distributions between these two configurations.
Because the center configuration has one extra electron in the center, and
a vacancy (hole) at the edge, this charge difference is positive near the 
center of the dot, and negative near the edge of the droplet.
Hence the crossing of orbitals leads to the redistribution of electrons from 
the edge (edge configuration) to the center (center configuration). 
The crossing, marked by squares in Fig.~\ref{dot0},
takes place at different magnetic fields $B_1$ for different particle numbers.
Varying the electron number allows us to trace this crossing of single 
particle levels.
This has been pointed out and calculated in a dot with low electron number and
strong kinetic energy quantization in Ref.~\onlinecite{arek secs fir}. 
Here we develop such a theory for large numbers of electrons.

With increasing magnetic field the Zeeman splitting becomes larger
than the kinetic energy $\Omega_-$ and the $\nu=2$ configuration is unstable 
against spin flip at the edge.\cite{hawrylak99}
In Fig.~\ref{dot0} this first spin flip is marked by empty circles for
even electron numbers, and by filled circles for odd electron numbers. 
We see that spin flips take place at different magnetic fields for even and 
odd electron numbers.
However, for a given parity of the electron number the magnetic field $B_2$ 
at which spin flip occurs does not depend on the number of electrons. 
The center-spin-flip line crosses the edge-spin-flip line at a critical 
particle number $N_c$.
Hence we might expect that the phase diagram of the $\nu=2$ droplet is finite,
i.e., the spin singlet phase exists only for a finite number of electrons.
The rest of the paper will be devoted to developing the understanding of 
how electron-electron interactions modify this single-particle picture.

\subsection{The many-particle Hamiltonian}

Denoting the creation (annihilation) operators for electrons 
in states $|m,n;\sigma\rangle$ by $c_{mn\sigma}^{+}(c_{mn\sigma})$,
the Hamiltonian of the interacting system can be written in second  
quantization as:
\begin{eqnarray}
H &=& \sum_{m,n,\sigma} 
\varepsilon_{mn\sigma} c^+_{mn\sigma} c_{mn\sigma} \nonumber \\
 &+ & \frac{1}{2}\sum_{m_1 n_1 m_2 n_2 m_3 n_3 m_4 n_4\sigma\sigma'} 
\langle m_1n_1,m_2n_2|V|m_3n_3,m_4n_4\rangle
c^+_{m_1n_1\sigma}c^+_{m_2n_2\sigma'}
 c_{m_3n_3\sigma'} c_{m_4n_4\sigma},
\label{ham}
\end{eqnarray}
where $\langle  m_1n_1,m_2n_2|V|m_3n_3m_4n_4\rangle$ are the
two-body Coulomb matrix elements defined in Ref.~\onlinecite{pawel ssc}. 
The Coulomb matrix elements conserve the angular momentum of the pair 
during the scattering process:
$m_1-n_1+m_2-n_2=m_3-n_3+m_4-n_4$.
They are measured in units of the exchange energy $E_{0}$:
$E_{0}={\cal R} \sqrt{ 2 \pi } a_{0} / l_{eff}$, where ${\cal R}$ is the 
effective Rydberg energy, $a_{0}$ is the effective Bohr radius, 
and $l_{eff}=l_{0}/(1+4\omega_0^2/\omega_c^2)^{1/4}$  is the effective 
magnetic length.
The Coulomb energy  increases with increasing magnetic field. 

\section{Quantum Hall droplet in the lowest Landau level approximation}

In this section we describe the properties of the $\nu=2$ droplet in the two 
lowest Landau level states neglecting mixing between them. 
This allows for a number of exact and analytical results which make
the physics transparent.

\subsection{Ground state of a QHD with 2N electrons}

In the lowest Landau level (LLL) the  ground state of the $\nu=2$ droplet is 
a product of two spin polarised droplets:
\begin{equation}
|GS(2N)\rangle = \prod_{m=0}^{N-1} c^+_{m,0\uparrow}  
\prod_{m=0}^{N-1} c^+_{m,0\downarrow} |0\rangle.
\label{hf wf lll}
\end{equation}
The total number of electrons $N_e=2N$, and the total angular momentum of 
the droplet, $R=2\sum_{m=0}^{N-1} m $, are good quantum numbers. 
In the LLL approximation this state is an exact ground state of the system.
It is useful to define electron self-energies $\Sigma(m,n,\sigma)$ at 
$\nu=2$ for a fixed number of electrons $2N$:
\begin{equation}
\Sigma(m,n,\sigma)=
\sum_{m'=0}^{N-1} (2 \langle  m,n;m',0|V|m',0;m,n\rangle 
     - \langle  m,n;m',0|V|m,n;m',0\rangle).
\end{equation}
This self-energy does not depend on spin, so in what follows the spin index 
will be dropped.
The total  energy $E_{GS}^{2N}=\langle GS(2N)|H|GS(2N)\rangle$
of the $\nu=2$ droplet is now given simply by a sum of energies of 
quasielectrons:
\begin{equation}
E_{GS}^{2N} =  \sum_{m=0}^{N-1} 2 \left(\Omega_{-}\left(m+\frac{1}{2}\right) + 
\frac{1}{2} \Omega_{+}\right)  +  \Sigma(m,0) .
\label{hf energy}
\end{equation}

The electrons are replaced by  quasiparticles (electrons dressed with 
interactions).

\subsection{Excitation spectrum and spectral functions of the QHD with 2N 
electrons}

Let us now analyze the charge and spin excitation spectrum of the $\nu=2$ 
droplet in the LLL.
The one-electron charge and spin excited states $| dm, m , \pm \rangle $ 
can be labeled by the increase
of angular momentum $dm$ with respect to the $\nu=2$ state.
We construct them by removing one electron from one of the occupied states 
$|m,\sigma\rangle$ and putting it onto one of the unoccupied states
$|m+dm,\sigma\rangle$:
\begin{equation}
| dm, m , \pm \rangle = 
{1\over\sqrt{2}}
[c^+_{m+dm,0,\downarrow} c_{m,0,\downarrow}
\pm
c^+_{m+dm,0,\uparrow} c_{m,0,\uparrow}]
|GS(2N)\rangle,
\end{equation}
where $+$ refers to spin singlet (charge excitation) and $-$ to spin triplet 
(spin excitation). 
To understand the difference between the two excitations we discuss the 
lowest angular momentum excited states $dm=+1$.
The spin singlet excitation  can be obtained by 
acting on the $\nu=2$ droplet with the center-of-mass operator 
$Q^+=\sum_{m,\sigma}  (m+1)^{1/2} c^+_{m+1,0,\sigma} c_{m,0,\sigma}$.
The energy of this excitation, measured from the energy of the 
$\nu=2$ state, equals exactly the kinetic energy $\Omega_-$, or the energy
needed to increase the angular momentum of one electron by one unit.
This is exactly what one finds for the $\nu=1$ state.
The spin triplet excitation on the other hand has the energy of the charge 
excitation minus twice the exchange energy across the Fermi level $N-1$,
i.e., $\Omega_- -2  \langle N-1;N|V|N-1;N\rangle$. 
Hence, the spin triplet excitations  have an energy lower than that of the
charge excitations.

The excitation spectrum can be probed by adding one electron to the droplet.
The probability of adding an electron to the orbital $|m,0\rangle$
with energy $E$ is given by the spectral function\cite{arek spectral}
$
A(m,E)=\sum_f \left|_f\langle 2N+1| c^+_{m,\downarrow} |GS(2N)\rangle\right|^2
\delta(E_f(2N+1)-E_0(2N)-E).
$
The electron probes all excited final states of the $2N+1$ droplet, a 
reflection of the excitation spectrum of the $2N$ droplet. 
In Fig.~\ref{spectral} we show an example of the calculated 
spectral function of a $2N=8$ electron droplet at $\nu=2$ obtained by 
exact diagonalisation techniques\cite{hawrylak99,arek spectral} for 
$\omega_c/\omega_0 = 0.2$.
The spectral function of the noninteracting system describes an addition of
an electron to empty states with energy spacing $\Omega_-$ and probablity of 
one.
In a Fermi liquid this noninteracting  picture would be only slightly modified
by interactions.
However, in the quantum dot the quasiparticle picture breaks down already 
at the first excited state, 
and for the energy $\sim 3\Omega_-$ the spectral function is already almost
zero.
Moreover, we see that even though there are many more states than in the
noninteracting picture, these states form bunches which leads to a discrete 
density of states.
The bunching and the separation of bunches is controlled by Coulomb 
interactions rather than by the single particle energy levels. 

\subsection{Spin flip excitation spectrum of QHD with 2N electrons}

Let us now turn to a detailed analysis of the spin flip excitations of
the $\nu=2$ droplet. 
We start with the center configuration.
It involves  an electron transfer from the highest occupied orbital of the 
lowest Landau level to the lowest energy state of the second Landau level 
accompanied by a spin flip:
$
|C(2N)\rangle=
c^+_{0,1,\downarrow} c_{N-1,0,\uparrow} |GS(2N)\rangle.
$
This state corresponds to two electrons with parallel spin (triplet
state), one in the center of the droplet and one at the edge.
There is only one state with this angular momentum and spin in the subspace 
of one-pair excitations on two Landau levels, so this is an exact eigenstate 
of the Hamiltonian. The two electrons involved cannot be distinguished from the rest of
electrons. To account for all electrons it is better to think of a created 
electron as a quasi-electron and a removed electron as a quasi-hole. 
The energy of the electron-hole pair   
is given by a difference in the energy of a quasi-electron in the
center orbital and a quasi-hole at the edge of the droplet, plus their
attraction:
\begin{eqnarray}
E(C)^{2N} &=& E_{GS}^{2N}
+ \left( \Omega_+ + \Sigma (0,1) -\frac{1}{2} E_z \right)
- \left( (N-1) \Omega_- +\Sigma (N-1,0)+ \frac{1}{2} E_z \right) 
 \nonumber \\
&-&    \langle N-1,0;0,1|V|0,1;N-1,0\rangle.
\label{enc}
\end{eqnarray}

The energy of the quasi-electron and quasi-hole pair depends on the
total number of electrons in the $\nu=2$ droplet. 
This interpretation of the triplet state as a collective state differs from 
the two-electron model of Tarucha et al.\cite{tarucha00}
When the magnetic field $B$ is lowered beyond a critical value $B_1$, the energy
of the center configuration becomes lower than the energy of the 
$\nu=2$ configuration, the $\nu=2$ spin singlet droplet becomes unstable, 
and a spin triplet center configuration becomes the ground state.


We now turn to the breakdown of stability of the droplet due to spin flip at 
the edge.
The spin flip wavefunctions can be generated from the droplet at filling factor
$\nu=2$ by removing a spin up electron at $|N-1,0\rangle$
and creating a spin down electron at the first available state at the edge 
$|N,0\rangle$ (a spin exciton):
$
|E(2N)\rangle=c^+_{N,0,\downarrow} c_{N-1,0,\uparrow}|GS(2N)\rangle.
$
The process described here is equivalent to creating a hole (a missing spin-up
electron) below the Fermi level, and a spin down electron at the edge. 
Similarly to Eq.~(\ref{enc}), we can easily calculate the energy of this
configuration:
\begin{eqnarray}
E(E)^{2N}&=&E_{GS}^{2N} + \Omega_- - E_z \nonumber \\
&+&\Sigma (N,0)-\Sigma (N-1,0) - \langle N-1,0;N,0|V|N,0;N-1,0\rangle.
\label{ene}
\end{eqnarray}
As compared to the energy of the $\nu=2$ droplet, the energy of the edge 
spin-flip exciton is increased by the kinetic energy $\Omega_-$, and
 lowered by the Zeeman energy. 
The quasi-electron and the quasi-hole are dressed by their self-energies,
and they attract each other.
As the magnetic field is increased, the energy of the edge spin-flip 
configuration becomes lower than the energy of the $\nu=2$ spin singlet 
droplet, and the triplet edge configuration becomes the ground state.
The magnetic field at which this takes place will be referred to as $B_2$.
 The calculated  difference of self-energies across the
Fermi level is  much smaller  than the vertex correction, which  controls
the value of the magnetic field where the spin flip occurs. 
The vertex correction in turn depends on $N$, hence $B_2$ 
depends on the number of electrons.

\subsection{Phase diagram of QHD with 2N electrons}

We can now use Eqs.~(\ref{enc}),~(\ref{ene}) to determine the critical
magnetic fields $B_1$, $B_2$ as a function of the number of electrons $N_e=2N$
for different confinement energies $\omega_0$ (independent of $N$) and
Zeeman energy $E_z=0.02$ meV/T appropriate for GaAs. For the GaAs Zeeman energy
 and  e.g. $\omega_0=6$ meV , the Zeeman splitting 
of single-particle levels in Fig.~\ref{dot0} would  now be barely visible,
and no spin flips would occur in the experimentally accessible range of magnetic fields.  
Then, the model of noninteracting electrons would  not allow for edge spin 
flips,\cite{hawrylak99} however the center spin-flip transition would still be 
possible.
The inclusion of interactions modifies this picture.
The calculated phase diagrams are shown in Fig.~\ref{phaseLLL}.
We find that the stability of phases is strongly affected by $\omega_0$
to the extent that the $\nu=2$ phase does not seem to be stable for
$\omega_0 < 4$ meV.
For larger confinement energies the phase diagram qualitatively
resembles the noninteracting phase diagram of Fig.~\ref{dot0}.
Quantitatively, the main difference between these two approaches
is visible in the behavior of the $\nu=2$ - edge spin flip phase boundary,
independent on the number of electrons in the noninteracting picture, but
strongly affected by the size of the droplet of interacting electrons.
Also, in the single-particle picture the center spin flip - $\nu=2$ phase 
boundary has a parabolic shape (traces the kinetic energy of the
$|0,1,\downarrow\rangle$ orbital), which is not preserved in
Fig.~\ref{phaseLLL}.
The key observation common to both approaches is that for any 
$\omega_0 > 4$ meV the $\nu=2$ phase is stable only up to a critical number 
of electrons $N_c$.
This critical number of electrons increases with increasing $\omega_0$.
Thus, a large droplet is expected to consist of a bulk $\nu=2$ core and a 
spin polarised edge.

\subsection{Quasi-electron in QHD with $2N+1$ electrons}

In transport experiments one does not measure directly the energies of
configurations discussed above. 
The measured quantity is the chemical potential, i.e., the difference 
between the ground state energies of the $2N+1$- and $2N$-electron 
droplet.\cite{pawel secs}
Therefore now we must discuss the system of $2N+1$ electrons, where we 
consider the extra electron as a quasi-electron added to the $N_e=2N$ 
electron droplet.


The ground state configuration consists of the $\nu=2$ droplet
and an extra electron on the $|N,0\rangle$ orbital.
The  energy of this state is the energy of the $\nu=2$ droplet plus 
the energy of the quasi-electron:
\begin{equation}
E_{GS}^{2N+1} = E_{GS}^{2N}+ \left( N+\frac{1}{2}\right)\Omega_{-}  + 
\frac{1}{2}\Omega_{+} - \frac{1}{2} E_z  + \Sigma (N,0),
\label{qe edge energy}
\end{equation}
which includes the self-energy evaluated for the $2N$ electron droplet.


The  center configuration is obtained by transferring the quasi-electron
to the center orbital. 
The energy of this configuration is given by:
\begin{equation}
E(C)^{2N+1} = E_{GS}^{2N}+  \frac{1}{2}\Omega_{-} 
+ \left(1+\frac{1}{2}\right)\Omega_{+} - \frac{1}{2} E_z
 + \Sigma (0,1).
\label{qe center energy}
\end{equation}


The first-spin-flip state can be generated from the droplet at filling factor
$\nu=2$ and a quasi-electron at the edge by removing a spin up electron at 
$|N-1,0\rangle$  and creating a spin down electron at the first available
state at the edge, i.e., $|N+1,0\rangle$.
This is the $\nu=2$ configuration with one hole and two
additional spin down electrons at the edge,
 a spin-flip trion, whose energy is given by:
\begin{eqnarray}
E(E)^{2N+1}&=&E_{GS}^{2N}+ 
\left[
  \left(N+\frac{1}{2}\right)\Omega_{-}  + \frac{1}{2}\Omega_{+}
  - \frac{1}{2} E_z
  + \Sigma (N,0)
\right]  \nonumber \\
&+& \left[
\left(N+1+\frac{1}{2}\right)\Omega_{-}  + \frac{1}{2}\Omega_{+}
  - \frac{1}{2} E_z
  + \Sigma (N+1,0)
  \right] \nonumber \\
  &-&\left[
\left(N-1+\frac{1}{2}\right)\Omega_{-}  + \frac{1}{2}\Omega_{+}
  + \frac{1}{2} E_z
  + \Sigma (N-1,0)
  \right]
\nonumber \\
&-&  \langle N-1,0;N,0|V|N,0;N-1,0\rangle   - \langle N-1,0;N+1,0|V|N+1,0;N-1,0\rangle  \nonumber \\
&+&  \langle N+1,0;N,0|V|N,0;N+1,0\rangle-\langle N+1,0;N,0|V|N+1,0;N,0\rangle.
\label{odd spin flip}
\end{eqnarray}

The energy of the first spin flip configuration for odd electron numbers
is a sum of the energies of a quasi-electron at $m=N$, a
quasi-electron at $m=N+1$, a quasi-hole at $m=N-1$,
the attractive interaction of electrons with a quasi-hole, and the direct
 and exchange interaction among the two electrons of the trion.

In analogy to the $2N$ electron case, we expect to observe ranges of magnetic
field in which each of these three configurations becomes the ground state
of the QHD.

\subsection{Addition spectrum}

We can now calculate the addition spectrum (chemical potential) of the 
$2N$ electron droplet, defined as $\mu(2N)=E(2N+1)-E(2N)$. 
The chemical potential as a function of the magnetic field
 exhibits features corresponding to  changes
in the ground state energies of the $2N$ and of  $2N+1$ droplet.
Let us start with the $\nu=2$ $2N$ droplet and add an extra quasi-electron
at the edge. 
The addition  energy to do so, or the chemical potential in this magnetic field
range,
\begin{equation}
\mu_3(2N) = E_{GS}^{2N+1} - E_{GS}^{2N} =
 N\Omega_{-}  + \frac{1}{2}\Omega - \frac{1}{2} E_z  + \Sigma (N,0),
\end{equation}
is a sum of the  kinetic energy and self-energy at the edge of the droplet.
The kinetic energy contribution {\em increases}  with decreasing magnetic field.
At a critical value
$B_1^*$ of the magnetic field the quasi-electron transfers from the edge to the center of the dot (transition
in the $2N+1$ droplet), while the $2N$ electron droplet remains stable.
In this new configuration  the addition energy
\begin{equation}
\mu_2(2N) =  E(C)^{2N+1} - E_{GS}^{2N} =
\Omega_{+} + \frac{1}{2}\Omega - \frac{1}{2} E_z  + \Sigma (0,1),
\end{equation}
is a sum of the kinetic  and the self-energy
$\Sigma (0,1)$ of the quasi-electron
in the center of the droplet.
The energy to add a quasi-electron to the center is proportional to
$\Omega_{+}$ and {\em decreases} with decreasing magnetic field. 
Therefore the chemical potential has an upward cusp at $B=B_1^*$.

When the  magnetic field is lowered below $B_1$, the $2N$ electron droplet 
becomes unstable and a spin triplet magneto-exciton center configuration is 
formed.
The magneto-exciton consists of an electron in the center and a hole at the 
edge.
The electron in the center cancels out in the additon spectrum and the chemical
potential now measures the energy of a "dressed" quasi-hole at the edge of 
the $2N$ droplet, the "dressing" coming from  final state correction in the 
form of electron-hole attraction:
\begin{eqnarray}
\mu_1(2N) &=& E(C)^{2N+1} - E(C)^{2N} \nonumber\\
&=&\frac{1}{2} \Omega + (N-1)\Omega_-  + \frac{1}{2} E_z 
+ \Sigma (N-1,0) + \langle N-1,0;0,1|V|0,1;N-1,0\rangle.
\end{eqnarray}
The kinetic energy of this hole {\em increases} with decreasing magnetic field.
 This tendency is somewhat compensated by the interactions.
Hence, at $B=B_1$ the addition spectrum exhibits a downward cusp.

Consider now the addition of one electron to an odd, $2N-1$-electron droplet.
In high magnetic fields the initial state consists of the $\nu=2$ droplet
and a hole on the  missing spin up state   at the edge.
The final state is the $2N$-electron $\nu=2$ droplet.
Thus the extra electron is added to the edge.
Its energy is proportional to $\Omega_-$ ( renormalized by
interactions which depend weakly on the magnetic field), therefore the chemical potential {\em increases} with the
decreasing magnetic field.
As we lower the magnetic field, the final state of $2N$ electrons undergoes
the first transition at $B=B_1$: the center configuration becomes stable.
The extra electron occupies the center orbital, the chemical
potential in this regime  {\em decreases} with decreasing magnetic field,
and the addition spectrum will exhibit a cusp pointing up.
Note that this cusp will be seen at exactly the same magnetic field
as the downward cusp described previously for the addition of an extra electron
to the even, $2N$-electron QHD.
Finally, both initial and final configurations contain a center electron.
Then, the addition of a spin down electron takes place at the edge, and
the chemical potential exhibits a downward cusp and begins to {\em increase} 
with decreasing magnetic field.

In Fig.~\ref{miu} we show the calculated chemical potentials for a droplet with
$N=16-21$ electrons, with charging energy subtracted. Each addition spectrum
contains a segment with energy increasing with increasing magnetic field.
We see that the spacing between consecutive pairs of addition curves 
oscillates between large and small.
For addition spectrum of a quasi-electron to a $2N$ electron droplet, 
the spacing corresponds to a difference between the energy to add a 
quasi-electron with spin down and the energy to add a quasi-hole with  
spin down (electron with spin up below the Fermi level).
In the interacting system, the electron and a quasi-hole are dressed by 
interactions. 
The energy difference, $\Delta_L$, is the difference in the chemical potential 
$\Delta_L=\mu_3(2N)-\mu_1(2N)$:
\begin{equation}
\Delta_L=\Omega_{-} - E_z + 
 \Sigma (N,0)-\Sigma (N-1,0) - \langle  N-1,0;0,1|V|0,1;N-1,0\rangle.
\end{equation}
Because this energy corresponds to a quasi-electron above the Fermi level and
a quasi-hole below, it includes the kinetic and self-energy difference
across the Fermi level, plus excitonic correction.
This is the final state interaction correction in the spectroscopy involving
transfer of electron from the edge into the center of the dot by decreasing
magnetic field.
The two spins are opposite and we also have to subtract the Zeeman energy. 
In a noninteracting system the spacing is proportional to 
$\Omega_{-} - E_z$.

Similar arguments can be used for the derivation of the other energy 
difference, $\Delta_S$, in the addition spectrum of the odd electron
number, $2N-1$ droplet:
\begin{equation}
\Delta_S= E_z + \langle  N-1,0;N-1,0|V|N-1,0;N-1,0\rangle
-\langle  N-1,0;0,1|V|0,1;N-1,0\rangle.
\end{equation}
The spacing for the odd electron droplet is proportional only to the Zeeman 
energy and excitonic corrections.
If interactions are weak, this spacing should be much smaller than the
spacing for even electron droplets. 
Therefore calculations predict a characteristic pattern in the addition
spectrum of the $\nu=2$ droplet related to the even or odd number of
electrons in the droplet. 
This pattern, measured experimentally,\cite{ciorga01b}
is expected to be valid only for electron numbers $N_e<N_c$.

\section{Landau level mixing, and exchange and correlation effects in QHD}

In the previous section we were able to derive a number of rigorous results by 
restricting Hilbert space to states derived from the two lowest Landau levels
without their mixing.
We now turn to include Landau level mixing and correlations.

\subsection{Hartree-Fock calculations in two lowest LL}

We start with the Hartree-Fock (HF) calculations, whose simplest example
involves only two Landau levels. 
We write the $\nu=2$ HF wave function as a product of variational wave 
functions in each spin channel $\sigma$ and angular momentum channel $m$:
\begin{equation}
|GS(2N)\rangle = \prod_{\sigma} \prod_{m=0}^{N-1} 
(a^*_{m,0,\sigma} c^+_{m,0,\sigma}
 + a^*_{m,1,\sigma} c^+_{m+1,1,\sigma})  |0\rangle.
\label{hf wf 2ll}
\end{equation}
The $N$ coefficients $a^*_{m,0,\sigma}$, $a^*_{m,1,\sigma}$, $m=0,1,..,N-1$ 
are determined by either minimizing the total energy subject to normalisation 
of the wave function or by solving the eigenvalue problem of the HF 
Hamiltonian. 
The HF Hamiltonian is defined by noting that the HF wave function allows us 
to define expectation values 
$\rho_{i,j}=\langle c^+_i c_j\rangle$ of the density matrix in terms of the 
coefficients $a^*_{m,0,\sigma}$, $a^*_{m,1,\sigma}$.
For example, 
$\langle c^+_{m,0,\sigma}c_{m,0,\sigma}\rangle=a^*_{m,0,\sigma}a_{m,0,\sigma}$,
$\langle c^+_{m+1,1,\sigma}c_{m+1,1,\sigma}\rangle
=a^*_{m,1,\sigma}a_{m,1,\sigma}$, 
are simply occupations of the HF level $0$ and the HF level $1$ in the angular
momentum channel $m$, while
$\langle c^+_{m+1,1,\sigma}c_{m,0,\sigma}\rangle 
=a^*_{m+1,1,\sigma}a_{m,0,\sigma}=p^*_{m,\sigma}$
are off-diagonal elements (polarisations) of the density matrix.
Replacing the two body term in Eq. (\ref{ham}) by a one body and mean-field 
term the HF Hamiltonian reads:
\begin{eqnarray}
H &=& \sum_{m,n,\sigma} 
\varepsilon_{m,n\sigma} c^+_{m,n\sigma} c_{m,n\sigma} \nonumber \\
 &+& \sum_{m_1 n_1 m_2 n_2 m_3 n_3 m_4 n_4 \sigma \sigma'} 
\langle m_1n_1,m_2n_2|V|m_3n_3,m_4n_4\rangle  \nonumber \\
& &\times
\left(c^+_{m_1n_1\sigma}\langle c^+_{m_2n_2\sigma'}c_{m_3n_3\sigma'}\rangle
c_{m_4n_4\sigma}-c^+_{m_1n_1\sigma}\langle 
c^+_{m_2n_2\sigma'}c_{m_4n_4\sigma} \rangle c_{m_3n_3\sigma'}\right).
\label{HFham}
\end{eqnarray}
The HF Hamiltonian can be further simplified by separating individual angular 
momentum and spin channels.
Neglecting the spin dependence, the effective Hamiltonian in a channel 
$l=m-n$ is simply a $2 \times 2$ matrix:
\begin{eqnarray}
H_{HF}=\left[
\begin{array}{cc}
  \varepsilon_{0}(l) + V_{0,0}  (l) &  V_{0,1}  (l)  \\
V_{0,1}  (l)  &  \varepsilon_{1}(l) + V_{1,1}
\end{array}
\right],
\end{eqnarray}
where $V_{i,j}(l)$ are self-consistent HF fields. 
For example, $V_{0,0}(l=m)$ reads
\begin{eqnarray}
&&V_{0,0}(l=m) =
\sum_{k} 
\langle m,0;k,0|V_{dx}|k,0;m,0\rangle a^*_{m,0} a_{m,0}+
\langle m,0;k+1,1|V_{dx}|k+1,1;m,0\rangle a^*_{m,1} a_{m,1}
\nonumber\\
&&+\langle m,0;k+1,1|V_{dx}|k,0;m,0\rangle a^*_{m,1} a_{m,0}+
\langle m,0;k,0|V_{dx}|k+1,1;m,0\rangle a^*_{m,0} a_{m,1},
\end{eqnarray}
with the direct and exchange matrix elements 
$\langle m\,0,k\,0|V_{dx}|k\,0,m\,0\rangle = 
2\langle m\,0,k\,0|V|k\,0,m\,0\rangle - \langle m\,0,k\,0|V|m\,0,k\,0\rangle$.
An example of the self-consistently calculated HF quasi-particle energies
$E_{HF}(l)$ for $N_e=12$, $B=3T$, and $\omega_0=6$ meV is shown in
Fig.~\ref{phaseHF}(a).
Circles denote quasiparticle energies calculated without Landau level 
mixing while bars show renormalised HF quasiparticle energies.
An overall behavior of quasiparticles is similar.
The self-consistent HF energies are however lower than the 
non-self-consistent ones.

A similar HF procedure can be applied to the center and edge spin flip
configurations, with the center configuration involving mixing of
$|m=0,n=1\rangle$ and $|m=1,n=2\rangle$ (third Landau level) states.
The resulting total energies can be compared to establish stability
ranges of different phases.
The resulting HF phase diagram for parameters identical to those in
Fig.~\ref{phaseLLL} is shown in Fig.~\ref{phaseHF}.
We see that the overall behavior of the phase diagram is similar without and
with Landau level mixing: there is a stable $\nu=2$ phase up to
a critical number of electrons, $N_c$, which depends on confinement energy.
However, the value of $N_c$ and the values of critical fields are
significantly different from those obtained in the LLL approximation.

\subsection{Influence of correlations - electron-hole pair excitations,
LSDA, and exact diagonalisation}

The center and edge spin flip processes described above involve
singlet-triplet transitions, hence they are driven by exchange interaction.
The HF approximation includes the exchange, but neglects electronic
correlations.
Correlations are known to counteract exchange, leading even to a collapse of
the Zeeman gap for QHDs in the $\nu<2$ regime.\cite{hawrylak99}
Therefore we attempt to include correlations into our model using three
different approaches:
(a) one electron-hole pair excitations from the self-consistent HF $\nu=2$ ground state,
(b) the SDFT/LSDA calculation over a broad range of electron numbers,
and (c) exact diagonalization for $N_e=6$ and $N_e=8$.

We start with one electron-hole pair excitation spectrum in the HF basis.
Our HF procedure is spin and space restricted, so the angular momentum and
the total spin $S_z$ are good quantum numbers.
We note that the  eigenvector corresponding to the lowest HF eigenvalue of
the self-consistent HF Hamiltonian determines coefficients
$a^*_{m,0,\sigma}, a^*_{m,1,\sigma}$ not only for occupied but also
for unoccupied angular momentum channels.
These coefficients allow us to replace the two operators
$c^+_{m,0,\sigma}, c^+_{m+1,1,\sigma}$ by a single operator 
$A^+_{m,0,\sigma}=a^*_{m,0,\sigma} c^+_{m,0,\sigma}
+a^*_{m,1,\sigma}c^+_{m+1,1,\sigma}$  
for all channels. 
The second self-consistent operator $A^+_{m,1,\sigma}$ can be naturally
constructed from the eigenvector corresponding to the second excited HF 
orbital.
We can therefore express our old creation operators $c^+$   in terms of new
operators $A^+$ as $A^+_{i,l}=\sum_j U_{i,j}(l) c^+_{l+j,j}$ ($i,j=0,1$)
in each angular momentum and spin channel $l\sigma$ and obtain the Hamiltonian 
in the new  basis: 
\begin{eqnarray}
H &=& \sum_{l,k,k',\sigma} 
t_{k,k'}(l) A^+_{k\,l,\sigma} A_{k',l,\sigma} \nonumber \\
&+& \frac{1}{2}\sum_{k_1 l_1 k_2 l_2 k_3 l_3 k_4 l_4\sigma\sigma'} 
\langle k_1l_1k_2l_2|V|k_3l_3k_4l_4\rangle A^+_{k_1l_1\sigma}
A^+_{k_2l_2\sigma'} A_{k_3l_3\sigma'} A_{k_4l_4\sigma},
\label{hamrot}
\end{eqnarray}
where $U$ is the transformation inverse to the one obtained 
from HF minimisation, and subsequent orthogonalisation procedure.

The renormalised Coulomb matrix elements are expressed now in terms of 
angular momentum $l$ and level index $k$ in each angular momentum channel as
\begin{eqnarray}
\langle k_1l_1k_2l_2|V|k_3l_3k_4l_4\rangle=
 \sum_{j_1j_2j_3j_4} &&
 \langle
 j_1+l_1\,k_1,j_2+l_2\,k_2|V|j_3+l_3\,k_3,j_4+l_4\,k_4\rangle
\nonumber\\
&&\times  
 U_{k_1,j_1}(l_1)U_{k_2,j_2}(l_2)U_{k_3,j_3}(l_3)U_{k_4,j_4}(l_4).
\end{eqnarray}
We see that the angular momentum structure of the two scattering particles 
is unchanged, but the distribution over Landau levels is changed.
Similar considerations hold for the renormalised kinetic energy matrix: 
\begin{equation}
t_{k,k'}(l)=\sum_j U_{kj}(l)U_{k'j}(l)\varepsilon_j(l),
\end{equation}
which is not diagonal in the HF basis.

For each angular momentum channel $L_{\nu=2}+\delta l$ we create
electron-hole pair triplet excitations from the $\nu=2$ ground state:
$A^+_{k,l+\delta l\uparrow} A_{0 l \downarrow} |\nu=2\rangle_{HF}$,
where $k=0$ corresponds to an excitation to the first Landau level,
possible only for $\delta l > 0$, and $k=1$ corresponds to an
excitation to the second Landau level, possible for any $\delta l > -N$
for a $2N$ electron system.
The correlated ground state wave function in this angular momentum channel can
be then written, in terms of pair excitations, as
$|\Psi(\delta l)\rangle= 
\sum_{k,l} D_{k,l} A^+_{k,l+\delta l\uparrow} A_{0 l \downarrow}
|\nu=2\rangle_{HF}$. 
The coefficients $D_{k,l}$ and energies of pair excitations are obtained by 
diagonalising the Hamiltonian matrix (\ref{hamrot}).

The pair excitation spectrum calculated for $2N=12$, $B=3$T, $\omega_0=6$ meV
is shown in Fig.~\ref{pairexcit} (a), together with HF quasiparticle energies.
The energy scale was shifted in such a way that the energy of the $\nu=2$
ground state is zero (Zeeman energy is neglected).
In the angular momentum channel $\delta l = -6$ there is only one state, because
only one pair excitation is possible (this is the center spin flip 
configuration). 
With the increase of $\delta l$ the number of inter-Landau level configurations increases.
The lowest excitation in each angular momentum channel  separates from a band
and forms a collective inter-Landau level spin flip mode.
Finally, at $\delta l = +1$ (the subspace of the edge spin flip), excitations
to the first Landau level become possible, and from now on two broad bands of 
states, separated by a gap, are visible. The lowest energy excitation in each angular momentum channel $\delta l\geq0$
also separates from a band
and forms a collective intra-Landau level edge spin flip mode.
In this approach, the center spin flip state is not renormalized by
correlations, but the edge spin flip configuration becomes correlated with
single electron-hole pair excitations to the second Landau level.

When the energy of spin flip excitations for any angular momentum
becomes lower than the
HF ground state, a transition in the droplet takes place.
We calculate these excitations over a broad range of electron numbers,
magnetic field, and confining energies, and obtain the phase diagram.
It is shown in Fig.~\ref{pairexcit} (b) (dash-dot lines).
For comparison  we also show the HF phase diagram obtained earlier
without  (dotted lines) and  with Landau level mixing  (solid lines).
As can be seen, both self-consistent HF and single pair excitation 
approximations predict
(a) the phase boundaries to occur for magnetic fields higher by $\sim 1$ T,
(b) the collapse of the $\nu=2$ phase to occur for smaller critical
number of electrons $N_c$
when compared to those obtained without Landau level mixing.

Let us now compare the HF (solid lines) and the pair excitation 
(dash-dot lines) approaches.
In both of them the low-field $\nu=2$ phase boundary (center spin flip)
is established by considering only one configuration.
Therefore the small discrepancy between the results cannot be attributed
to correlation effects, but rather to the difference between the two
approaches. In HF the energy of each configuration is found by separate
self-consistent procedures, in pair excitations only the $\nu=2$ configuration
is considered self-consistently, and the center spin flip is an excitation 
from $\nu=2$.
This effect also plays a role in the high field $\nu=2$ phase boundary,
but here the edge spin flip state is correlated with the configurations
involving occupation of the second Landau level.
Thus the influence of correlations  counteracts the exchange,
thereby broadening the $\nu=2$ phase stability range.

An alternative approach to include correlations for arbitrary electron number
is to include exchange and correlation in the SDFT/LSDA approach.
Its application proved insightful in the study of many ground state properties
of artificial atoms and molecules.\cite{ferconi vignale,austing99,koskinen97,hirose99,%
serra,steffens98a,steffens99,wensauer00,wensauer01}
We apply the local spin density approximation using Tanatar's and 
Ceperley's parametrization of the exchange-correlation 
potential.\cite{tanatar89} 
To remain consistent with previous approximations, we took only two
Landau levels as the variational space for the calculation.
The result is shown in Fig.~\ref{pairexcit} (b) (dashed lines).
The low-field $\nu=2$ stability edge calculated here does not differ 
substantially from that obtained previously.
However, due to correlations, the $\nu=2$ stability range is now increased by
 $~1.5$ T.
Moreover, the spin singlet remains stable up to $N_c=38$ electrons, a number 
twice  that predicted by HF approximation.
This stability is also due to the influence of correlations:
the exchange effects lower only the energy of spin-polarised states, while 
correlation effects provide a mechanism to decrease ground state energies for 
the spin singlet configuration as well.

On the basis of the comparison of several approaches we see that the stability of the spin singlet droplet
is a sensitive function of correlations. Different approximations give qualitatively similar results but differ
significantly in quantitative predictions.

In order to gauge the validity of different approximations, we calculated the phase diagram
using exact diagonalization techniques for electron numbers $N_e=6$ and $N_e=8$.
Details of this approach will be published elsewhere.\cite{exactdiag}
Here we only note that in this calculation the total angular momentum, 
total spin and total $S_z$ were resolved as good quantum numbers.
It reduced the Hilbert space for each set of quantum numbers to
a computable size, so that within the 2 Landau-level approximation
both exchange and correlations are treated exactly.
The results are shown in Fig.~\ref{pairexcit} (b) (full circles).
For $2N=6$ we see a stability region of the $\nu=2$ phase much broader than 
that from the pair excitation approximation, but the discrepancy
becomes much smaller for a larger number of electrons, where the 
self-consistent approach is expected to work better.
However, the discrepancy between the exact diagonalization and the SDFT/LSDA 
results does not seem to improve with the increase of the electron number.
To analyze this problem we calculated phase diagrams for $2N=6$ electrons 
using SDFT/LSDA and exact diagonalization including three Landau levels.
The results (not shown here) indicate that the spin density profiles
do not change dramatically with the increase of the number of Landau levels, 
and therefore the SDFT phase diagram remains similar to that in 
Fig.~\ref{pairexcit} (b).
On the other hand, the Hilbert space size in the exact diagonalization 
increases by  orders of magnitude, which  affects the
phase diagram, shifting it towards higher magnetic fields.
Approximate calculations for 8 Landau levels reveal a good agreement
between SDFT and exact diagonalization., and we rely on predictions of
SDFT for large electron numbers in what follows.


\section{CB amplitude reversal}

Up to now we discussed the $\nu=2$ line as the set of features in the position of CB lines
reflecting the transfer of one electron from the center to the edge of the QHD.
We assumed the electron number to be small enough, so that this transition involved the
spin singlet configuration of the even-electron QHD.
We established that the existence of pairs of cusps seen in the CB trace was due
to a different energy dependence of the center and edge orbitals on the magnetic field.
We now turn to describing the behaviour of the droplet in the vicinity of the
critical electron number $N_c$, where the spin singlet configuration ceases to 
be the ground state.
As can be seen in Figs.~\ref{phaseHF},~\ref{pairexcit} (b), beyond $N_c$ the
$\nu=2$ line is continued by another phase boundary, separating the center spin
polarised and the edge spin polarised configurations.
The transition between these two phases involves the transfer of one electron 
from
the center to the edge of the QHD, but this time the final state is a spin triplet.
Because of that the continuation of the $\nu=2$ line in the CB trace will
exhibit a pair of cusps similar to that observed for a the electron number $2N<N_c$.
Thus the CB peak {\em position} spectroscopy is not expected to reveal any indication of 
the collapse of the spin singlet phase.

The tunneling spectroscopy conducted in the regime of spin polarised 
injection/detection\cite{ciorga01b} allows, however, to extract information also from
the CB peak {\em amplitude}.
In this regime the energy separation of two different spin species in the
two-dimensional elecron gas (2DEG) making up the source and drain reservoirs is converted
to the spatial separation of electrons with different spin orientations
at the edge of the 2DEG.\cite{ciorga01b,hawrylak99}
This spatial separation at the edge drastically influences the tunneling probability of 
electrons with different spin to and from a quantum dot.
Now the electrons injected into the dot are predominantly spin down.
Moreover, the tunneling matrix elements are dominated by the overlap of the wave
functions of the dot and the reservoirs.
Therefore the tunneling event is allowed if the extra spin down
electron is added to (or removed from) the edge of the QHD.
If the extra electron is spin up, the current amplitude is expected to be much smaller
(spin blockade), and if we add a spin down electron to the center, the current amplitude 
is small due to small wave function overlap.
This spin-resolved tunneling should then be sensitive to the collapse of the 
spin singlet phase, therefore we now focus on the understanding of CB amplitude
patterns both below and above the critical electron number.

\subsection{Configurations and corresponding amplitudes}

Let us first consider the system of noninteracting electrons with artificially
enhanced Zeeman energy (Fig.~\ref{dot0}).
The collapse of the $\nu=2$ phase takes place when the energy of the center orbital
crosses the first edge spin flip line denoted by full circles, i.e., at the magnetic 
field $7.8$T, for $N_c=14$ electrons. 

The configurations adjacent to the $\nu=2$ line in the regime of small number of
electrons ($2N<N_c$) are schematically shown in Fig.~\ref{confnormal} (a) for
four neighboring electron numbers, $2N$ through $2N+3$.
Let us analyze systematically the expected amplitude pattern starting with the
addition of one electron to the $2N$-electron QHD.
For  low magnetic field we add the electron to the center configuration to obtain
the center configuration as a final state.
The extra electron is added at the edge, but its spin is up, and the spin blockade 
causes the tunneling current amplitude to be small.
As we increase the magnetic field, the $2N$-electron QHD undergoes a transition to the
spin singlet.
The extra electron must be added to the center, but due to small overlap of wave
functions the tunneling amplitude is very small.
Finally, the $2N+1$-electron QHD undergoes a center-edge transition.
In this range of magnetic field the extra electron spin down is added to the edge,
so we expect the tunneling current to have a high amplitude.

Let us now add one electron to the $2N+1$-electron system.
For low magnetic fields both $2N+1$- and $2N+2$-electron systems are in the respective
center configurations.
They differ by one spin down electron  at the edge, so the tunneling current amplitude
is high.
For intermediate magnetic fields the $2N+1$-electron droplet exhibits the electron
transfer to the edge and we must add the spin down electron to the center, which
causes the amplitude to be small.
For higher magnetic fields the $2N+2$-electron droplet becomes a spin singlet,
the extra electron must be added at the edge, but with spin up.
The spin blockade causes the amplitude to be low.
The traces of CB peaks corresponding to these transitions are shown in
Fig.~\ref{confnormal} (b), with thicker lines marking the sections where a high
amplitude is expected.

If the number of electrons $2N>N_c$, the spin singlet phase is no longer stable, and
the configurations adjacent to the $\nu=2$ line are slightly different than those in the
normal regime (see Fig.~\ref{confreverse}).
The center configuration for odd electron numbers, instead of being a singlet with
one extra electron in the center, acquires a spin polarised edge.
Similarly, the edge configuration for even electron numbers, previously a spin singlet,
becomes edge spin-polarised.
These new configurations strongly influence the addition amplitudes.
Upon careful consideration of spin blockade and overlap conditions, we find that
now the amplitude pattern is opposite to that observed in the normal regime:
 the amplitude is high on the low-magnetic field side of the $\nu=2$ line
when we add an extra electron to the even-electron system.
This is summarized in the addition spectrum shown in Fig.~\ref{confreverse} (b).

As we can see, the collapse of the $\nu=2$ phase manifests itself in the CB 
spectroscopy by the reversal of the CB current amplitude patterns, as observed in
recent experiments.\cite{ciorga01b}
However, when we increase the magnetic field and the number of electrons further,
we encounter the second edge spin flip line, denoted in Fig.~\ref{dot0} by
full circles. 
In this regime the configurations adjacent to the $\nu=2$ line will change,
resulting in yet another modification of the amplitude pattern.
This increasing complexity forces us to go beyond a qualitative model and examine the
influence of interactions on our system.

\subsection{Mean-field calculations}

We choose two complementary approaches discussed previously: the
HF approach in two Landau level approximation and the SDFT/LSDA approach.
This choice permits us to examine the influence of direct and exchange interactions
and of correlations separately.

The HF analysis is carried out in the way described before, but now it is extended
to configurations of both odd and even electron numbers with higher polarised edge.
We used the parameters: $\omega_0=6$ meV and $E_z=0.02$ meV/T.
The resulting phase diagrams are shown in Fig.~\ref{phase2ll}.
For even electron numbers we observe the breakdown of the $\nu=2$ phase at $2N=20$
electrons.
However, the higher polarised phases do not align favourably for the system to exibit
the reversal of amplitudes.
In order to observe it, we need to have the center and edge configurations both
with total $S_z=-1$ neighboring  the $\nu=2$ line.
However, as can be seen from Fig.~\ref{phase2ll} (a), the center configuration
with $S_z=-1$ ceases to be stable in favour of the higher spin-polarised center 
configuration with $S_z=-2$ right at the magnetic field and the electron number 
for which the spin singlet collapses.
Compared to the single-particle picture this transition is shifted to much lower
magnetic fields.
Therefore it can be argued that the favourable alignment of phases takes place
just for one electron number, $2N=20$.
As for the odd electron numbers (Fig.~\ref{phase2ll} (b)), the phases are aligned 
favourably for $2N+1=19$ and 21 electrons.

So, the characteristic reversal of amplitudes predicted in the single-particle 
picture will be now observed only in transitions from 19- to
20-electron and from 20- to 21-electron QHD.
The reason of this behaviour of the system is the overestimation of the 
exchange interaction by the HF approach.
Because of that, the spin polarised configuration with total $S_z=-2$ will have 
artificially lowered energy with respect to the configuration with total $S_z=-1$ 
and the transition between these two configurations will occur too early.
We established earlier that correlations can counteract these artificially enhanced
exchange interactions.
Therefore we will now generate the analogous phase diagrams using the SDFT/LSDA 
approximation.

In what follows, in order to provide a sufficient variational space for the DFT 
algorithm, we took into account a single-particle basis set with quantum numbers 
$n=0,\dots,9$ and $m=0,\dots,59$, i.e., 10 Landau levels.
To make contact with experiment,  the confinement energy $\omega_0=1$ meV, close
to the value observed experimentally\cite{ciorga01b} has been used.
We carried out two calculations: for the normal GaAs Zeeman energy $E_z=0.02$
meV/T (Fig.~\ref{phasesdft1}), and for the Zeeman energy artificially enhanced
by a factor of two (Fig.~\ref{phasesdft2}). 
Let us focus on the odd electron numbers first.
Compared to the HF calculation, the range of favourable alignment of phases
has been vastly extended - to 5 electron numbers (43 through 51) for normal Zeeman
energy, and even 7 electron numbers (35 through 47) for enhanced Zeeman energy.
For even electron numbers we observe the breakdown of the spin singlet phase for
$N_c=48$ and 38 electrons with normal and doubled Zeeman energy, respectively.
However, in Fig.~\ref{phasesdft1} for normal Zeeman energy the favourable
alignment of phases is again seen only for one electron number, $2N=50$,
so the amplitude reversal is predicted only for QHD transitions from 49 to
50 electrons and from 50 to 51 electrons.
The situation is improved only by enhancing the Zeeman energy (Fig.~\ref{phasesdft2} (a)),
which shifts down the energies of spin polarised configurations without changing
the energy of the spin singlet.
As a result the collapse of the $\nu=2$ phase takes place for lower magnetic fields and
lower electron number, and the spin flips in question are finally disaligned.
We then find a stable reversal of the amplitude modulation for QHDs with 38 to 43 electrons. 
Note that the sole purpose of enhancing the Zeeman energy was to favorize spin polarised
configurations over the spin singlet.
This goal could be also attained by including more exchange interaction, 
perhaps slightly underestimated in the SDFT.
Therefore the reversal of amplitude patterns is very sensitive to the balance between
exchange and correlations.

\section{Conclusions and outlook}

We investigated the physics of the $\nu=2$  quantum Hall droplet as a
function of the electron number, confinement energy,  and the magnetic field.
We considered the spin singlet configuration and compared its energy to the
energies of configurations with increasing spin polarization, both for even
and odd electron numbers.
We found that there exists a critical number of electrons $N_c$, beyond which
the spin singlet phase ceases to be stable at any magnetic field.
Then, by considering the energies of configurations adjacent to the $\nu=2$
line for the $2N$ system, and comparing them to those of the $2N+1$ system,
we established the characteristic  shape of the addition spectra,
due to center-edge electronic transitions in the initial or final state.
We studied the same configurations in different  approaches (HF,
electron-hole pair excitation spectra, and SDFT/LSDA) to take into
account the electron-electron interactions. 
All our calculations clearly show the breakdown of the $\nu=2$ quantum Hall
droplet as a GS configuration with increasing number of electrons
and magnetic field.
However, depending on the method, this breakdown occurs in different regimes. 
The Hartree-Fock calculations, treating the exchange energy but neglecting 
correlations, uncover the basic renormalization effects of the interaction.
It acts effectively as an enhanced Zeeman energy, lowering the energies of
spin polarised states compared to the noninteracting picture, and
causing the spin flip transitions to occur for experimentally observed
magnetic fields (below 10 T),
and the collapse of the spin singlet phase for a critical number of electrons
of order of $N_c\sim 30$.
The inclusion of correlations, carried out by employing the electron-hole
pair excitation approach, the SDFT/LSDA approximation, and exact diagonalisation techniques
leads to a broadening
of the stability range of the spin singlet phase both in magnetic field and in
electron number. 
The reason for this is that unlike the exchange, correlations
lower the energies of spin singlet as well as spin polarised configurations.

Then we turned to predicting the signature of the collapse of the spin singlet
phase in the addition spectrum.
We found that it will not be revealed in the Coulomb blockade peak position, but
will be manifested by a reversal of the CB current amplitude patterns.
This reversal of amplitudes is almost not recovered upon the inclusion of 
interactions if the  approximation used overestimates exchange or correlations.
However, if a proper balance between them is found (e.g., by using an
enhanced Zeeman energy), the reversal of amplitude pattern is obtained for
a range of  electron numbers.
These theoretical predictions regarding both CB peak position and amplitude 
spectra are observed experimentally.\cite{ciorga01b}

A theoretically and experimentally interesting problem not discussed here is the
multiple reversal of the amplitude pattern as suggested by consideration of the
noninteracting system. 
Unfortunately, the experimental investigations have not yet been carried out 
up to high enough electron number to observe it.

\section{Acknowledgement}
We thank Jordan Kyriakidis, Andy Sachrajda, Mariusz Ciorga, and Michel Pioro-Ladriere
for helpful discussions.
A.\ W.\ acknowledges  the German Academic Exchange Service (Grant no. D/00/05486)
and Institute for Microstructural Sciences for financial support.

\begin{figure}[ht]
\caption{\label{dot0}
(a) Magnetic field evolution of single-particle energies for
$\Omega_0=6$ meV and artificially enhanced Zeeman energy $E_z=0.15$ meV/T.
Circles denote the edge spin flip of
a droplet with even (empty) and odd (full) number of electrons.
Squares denote the center spin flip.
(b) Configuration of noninteracting electrons corresponding to the
$\nu=2$ spin-singled quantum Hall droplet.
}
\end{figure}

\begin{figure}[ht]
\caption{\label{wf}
(a) Probability distribution of single particle orbitals $|0,1\rangle$
(dot center) and $|9,0\rangle$ (dot edge).
(b) Charge distribution in the quantum Hall droplet in the center
configuration, the $\nu=2$ phase, and difference in charge
distribution of the droplet due to the edge-center transition.
}
\end{figure}

\begin{figure}[ht]
\caption{\label{spectral}
Spectral function of the $\nu=2$ droplet with N=8 in the lowest Landau
level approximation.}
\end{figure}

\begin{figure}[ht]
\caption{\label{phaseLLL}
Phase diagram of the even electron droplet as a function of 
electron number and the magnetic field in the lowest Landau level
approximation and for different confinement energies.}
\end{figure}

\begin{figure}[ht]
\caption{\label{miu}
(a) Schematic view of the chemical potentials for odd and even
number of electrons indicating the energy differences 
$\Delta_L$ and $\Delta_S$ discussed in the text.
(b) Addition spectrum of a quantum dot in a spin blockade 
experiment.\cite{ciorga01b}}
\end{figure}

\begin{figure}[ht]
\caption{\label{phaseHF}
(a) Hartree-Fock quasiparticle energies for a dot with 8 electrons,
confinement energy $\omega_0=6$meV and in the magnetic field $3$T; the
circles (bars) [symbols] show the energies derived without (with) Landau level
mixing. (b) Phase diagram of the even electron droplet as a function of the
electron number and magnetic field in the Hartree-Fock approximation. }
\end{figure}

\begin{figure}[ht]
\caption{\label{pairexcit}
(a) Pair excitation spectrum from the $\nu=2$ droplet for 12
electrons, $B=3$T, $\omega_0=6$ meV for different angular momentum channels. 
Black circles denote the Hartree-Fock quasiparticle energies, bars denote 
correlated triplet ground and excited states.  
(b) Phase diagrams obtained without Landau level mixing (dotted lines),
using Hartree-Fock approximation (solid lines), from single electron-hole pair
excitation spectra (dot-dash lines), using SDFT/LSDA approach (dashed lines) 
and exact diagonalization (full circles).}
\end{figure}

\begin{figure}[ht]
\caption{\label{confnormal}
Ground state configurations (a) and schematic addition and amplitude spectrum 
(b) in the vicinity of the $\nu=2$ line for QHD with small number of 
electrons.}
\end{figure}

\begin{figure}[ht]
\caption{\label{confreverse}
Ground state configurations (a) and schematic addition and amplitude spectrum 
(b) in the vicinity of the $\nu=2$ line for QHD with large number of 
electrons.}
\end{figure}

\begin{figure}[ht]
\caption{\label{phase2ll}
Phase diagram of the interacting system with GaAs
  Zeeman energy for even (a) and odd (b) electron numbers 
  in the HF approximation. The center and edge configurations are separated by 
  the thick line, the states for even and odd electron numbers can be 
  identified by their spin quantum number. }
\end{figure}

\begin{figure}[ht]
\caption{\label{phasesdft1}
Phase diagram of the interacting system with GaAs
  Zeeman energy for even (a) and odd (b) electron numbers in the 
  SDFT/LSDA approximation including 10 Landau levels. The center and edge 
  configurations are separated by the thick line, the states for even and odd
  electron numbers can be identified by their spin quantum number. }
\end{figure}

\begin{figure}[ht]
\caption{\label{phasesdft2}
Phase diagram of the interacting system with doubled
  Zeeman energy for even (a) and odd (b) electron numbers in the 
  SDFT/LSDA approximation including 10 Landau levels. The
  center and edge configurations are separated by 
  the thick line, the states for even and odd
  electron numbers can be identified by their spin quantum number.
  The shaded areas mark the regime which shows a reversal of
  amplitudes. }
\end{figure}


\begin{thebibliography}{}

\bibitem{dot reviews}
For reviews and references see
L. Jacak, P. Hawrylak, and A. W\'ojs,
Quantum Dots, Springer Verlag Berlin, 1998;
L. P. Kouwenhoven, C. M. Marcus, P. McEuen, S. Tarucha, R.
Westervelt and N. S. Wingreen, Electron Transport in Quantum Dots in {\it %
Mesoscopic Electron Transport}, edited by L. L. Sohn, L. P. Kouwenhoven and
G. Schon (Kluwer, Series E 345, 1997);
R. C. Ashoori, Nature {\bf 379},413 (1996);
M. Kastner, Physics Today, {\bf 46}, 24 (1993);
T. Chakraborty, Comments in Cond.Matter Physics {\bf 16},35(1992);


\bibitem{ciorga00}
M. Ciorga, A.S. Sachrajda, P. Hawrylak, C. Gould,
P. Zawadzki, Y. Feng, Z. Wasilewski, Phys. Rev. B {\bf 61}, R16315 (2000).

\bibitem{tarucha00}
S. Tarucha, D.G. Austing, Y. Tokura, W.G. van der Wiel, L.P. Kouvehoven,
Phys. Rev. Lett. {\bf 84}, 2485 (2000).

\bibitem{hawrylak99}
P. Hawrylak, C. Gould, A.S. Sachrajda, Y. Feng, Z. Wasilewski,
Phys. Rev. B {\bf 59}, 2801 (1999).

\bibitem{austing99}
D.G. Austing, S. Sasaki, S. Tarucha, S.M. Reimann, M. Koskinen, M. Manninen,
Phys. Rev. B {\bf 60}, 11514 (1999).

\bibitem{tarucha96}
S. Tarucha, D.G. Austing, T. Honda, R. J. van der Haage, and
L. P. Kouwenhoven, Phys. Rev. Lett. {\bf 77}, 3613 (1996).

\bibitem{ray secs}
R.C. Ashoori, H.L. Stormer, J.S. Weiner, L.N. Pfeiffer,
K.W. Baldwin, and K.W. West, Phys. Rev. Lett. {\bf 71}, 613 (1993).

\bibitem{mceuen}
P.L. McEuen, E.B. Foxman, U. Meirav, M.A. Kastner, Y. Meir,
N.S. Wingreen, and S.J. Wind, Phys. Rev. Lett. {\bf 66}, 1926 (1991);
P.L. McEuen, E.B. Foxman, J.M. Kinaret, U. Meirav, M.A. Kastner,
N.S. Wingreen, and S.J. Wind, Phys. Rev. B {\bf 45}, 11 419 (1992).


\bibitem{klein}
O. Klein, S. de Chamon, D. Tang, D. M. Abusch-Magder, U. Meirav,
X.-G. Wen, M. A. Kastner, and S. J. Wind,
Phys. Rev. Lett. {\bf 74}, 785 (1995).


\bibitem{osterkamp}
T. H. Oosterkamp, J. W. Janssen, L. P.Kouwenhoven, D. G. Austing, T.
Honda, and S. Tarucha,Phys. Rev. Lett. {\bf 82}, 2931 (1999)

\bibitem{sachrajda}
A.S. Sachrajda, R.P. Taylor, C. Dharma-wardana, P. Zawadzki, J.A. Adams,
and P. T. Coleridge, Phys. Rev. B {\bf 47}, 6811 (1993).

\bibitem{marcus}
D.R. Stewart, D. Sprinzak, C.M. Marcus,
C.I. Duruoz, J.S. Harris, Science {\bf 278}, 1784 (1997).

\bibitem{weis}
J. Weis, R.J. Haug, K. von Klitzing, K. Ploog,
Phys. Rev. Lett. {\bf 71}, 4019 (1993).

\bibitem{blick}
R.H. Blick, D. Pfannkuche, R.J. Haug, K. von Klitzing, and K. Eberl
Phys. Rev. Lett. {\bf 80}, 4032 (1998).



\bibitem{pawel secs}
P. Hawrylak, Phys. Rev. Lett. {\bf 71}, 3347 (1993).


\bibitem{allan dot}
A. H. MacDonald, S. R. Eric Yang, and M. D. Johnson,
Aust. J. Phys. {\bf46}, 345 (1993).


\bibitem{eric allan mike}
S.R. Eric Yang, A.H. MacDonald, and M.D. Johnson,
Phys. Rev. Lett. {\bf 71}, 3194 (1993).



\bibitem{wen}
X. G. Wen, Phys. Rev. B, {\bf41}, 12 838 (1990);
C. de Chamon and X.-G. Wen, Phys. Rev.{\bf B49}, 8227 (1994).



\bibitem{oaknin}
J.H. Oaknin, L. Martin-Moreno, and C. Tejedor,
Phys. Rev. B {\bf 54}, 16 850 (1996);
J.J. Palacios, L. Martin-Moreno, G. Chiappe, E. Louis,
and C. Tejedor, Phys. Rev. B {\bf 50}, 5760 (1994).

\bibitem{imamura}
H. Imamura, H. Aoki, P.A. Maksym,
Phys. Rev. B {\bf 57}, R4257 (1998).

\bibitem{pawel dot imp}
P. Hawrylak, Phys. Rev. B {\bf 51}, 17 708 (1995).

\bibitem{arek spectral}
A. Wojs and P. Hawrylak, Phys. Rev. B {\bf 56}, 13227 (1997).


\bibitem{ciorga01}
M. Ciorga, A. Wensauer, M. Pioro-Ladriere, M. Korkusinski,
J. Kyriakidis, A.S. Sachrajda, P. Hawrylak, submitted to Phys.Rev.Letterse.

\bibitem{arek secs fir}
A. Wojs and P. Hawrylak, Phys. Rev. B {\bf 53}, 10 841 (1996).

\bibitem{pawel ssc}
P. Hawrylak, Solid State Commun. {\bf 88}, 475 (1993).

\bibitem{ciorga01b}
M. Ciorga, A.S. Sachrajda, P. Hawrylak, C. Gould, P. Zawadzki,
Y. Feng, Z. Wasilewski, Physica E {\bf 11}, 35 (2001);
A.S. Sachrajda, P. Hawrylak, M. Ciorga, C. Gould, and P. Zawadzki,
Physica E {\bf 10}, 493 (2001).



\bibitem{ferconi vignale}
M. Ferconi and G. Vignale, Phys. Rev. B {\bf 56}, 12108 (1997).


\bibitem{koskinen97}
M. Koskinen, M. Manninen, S.M. Reimann,
Phys. Rev. Lett. {\bf 79}, 1389 (1997).

\bibitem{hirose99}
K. Hirose, N.S. Wingreen, Phys. Rev. B {\bf 59}, 4604 (1999).


\bibitem{serra}
A. Puente and Ll Serra, Phys. Rev. Lett. {\bf 83}, 3266 (1999).


\bibitem{steffens98a}
O. Steffens, U. R\"ossler, and M. Suhrke, Europhys. Lett. {\bf 42},529 (1998).

\bibitem{steffens99}
O. Steffens, M. Suhrke, Phys. Rev. Lett. {\bf 82}, 3891 (1999).

\bibitem{wensauer00}
A. Wensauer, O. Steffens, M. Suhrke, U. R\"ossler,
Phys. Rev. B {\bf 62}, 2605 (2000).

\bibitem{wensauer01}
A. Wensauer, J. Kainz, M. Suhrke, U. R\"ossler,
Phys. Stat. Sol. (B) {\bf 224}, 675, (2001).

\bibitem{tanatar89}
B. Tanatar and D.M. Ceperley, Phys. Rev. B {\bf 39}, 5005 (1989).

\bibitem{exactdiag}
A. Wensauer, M. Korkusinski and P. Hawrylak, in preparation.

\end{thebibliography}
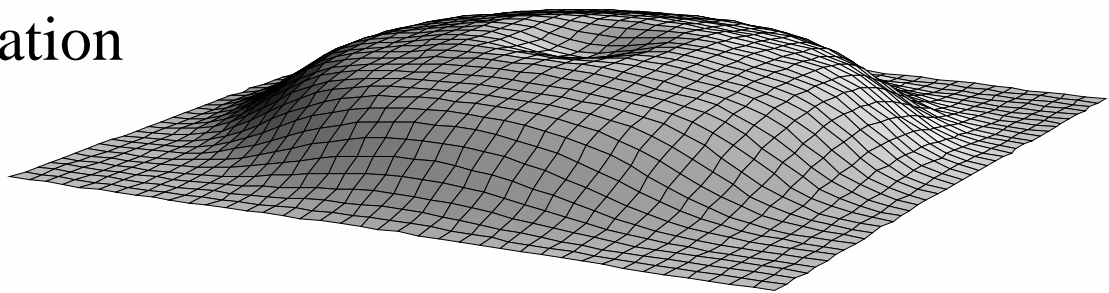
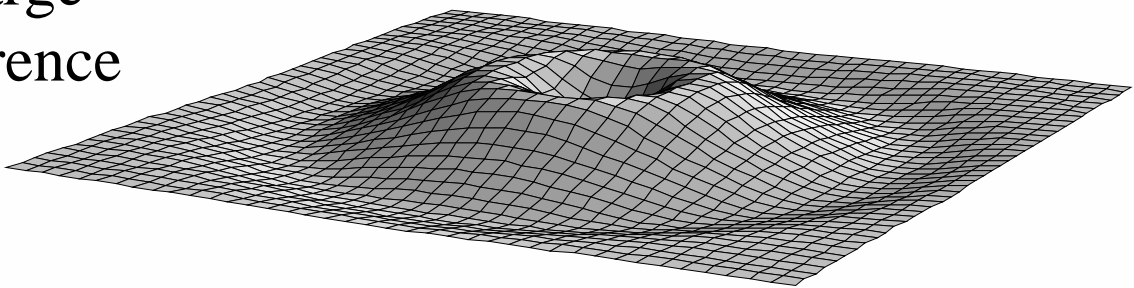
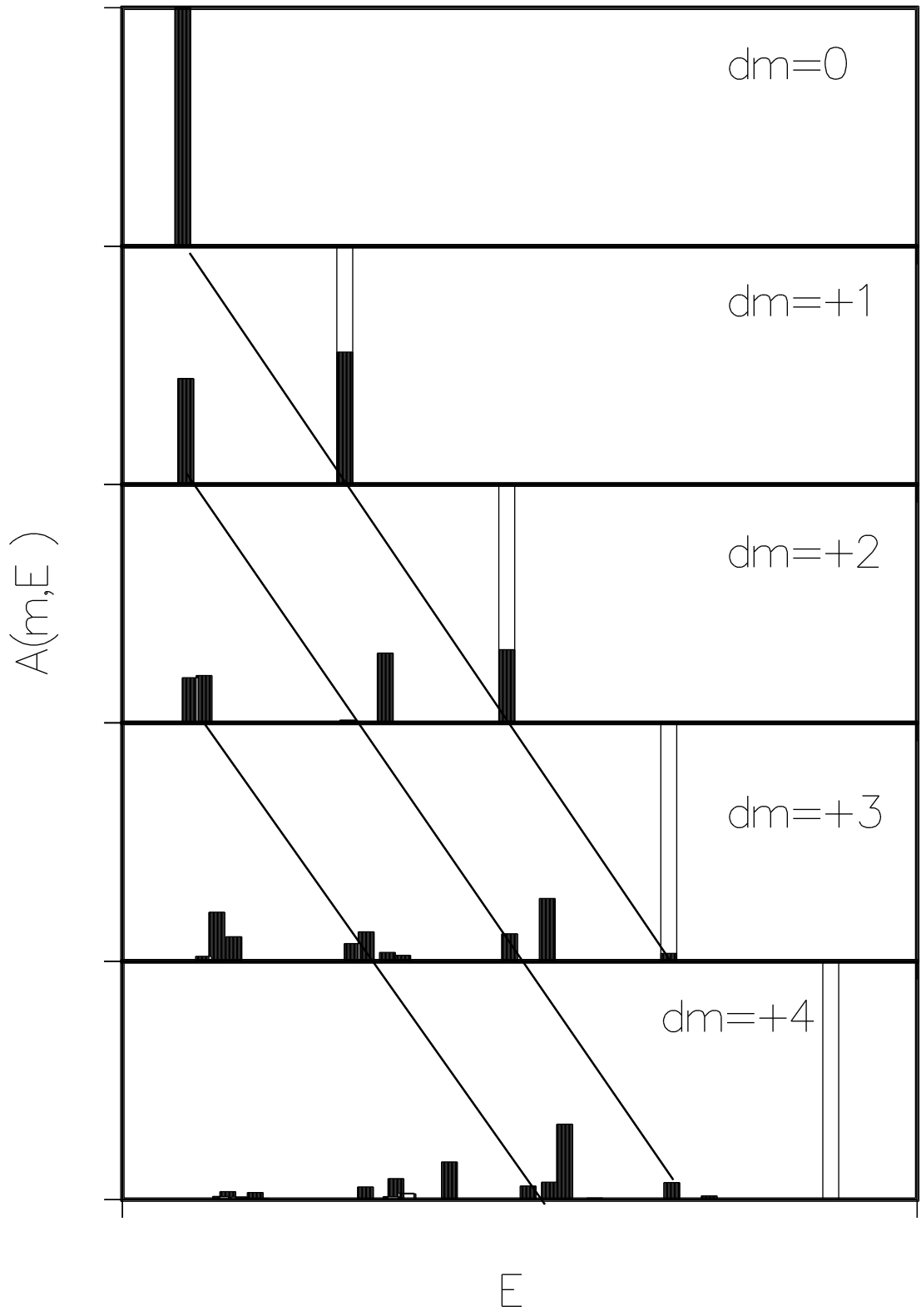
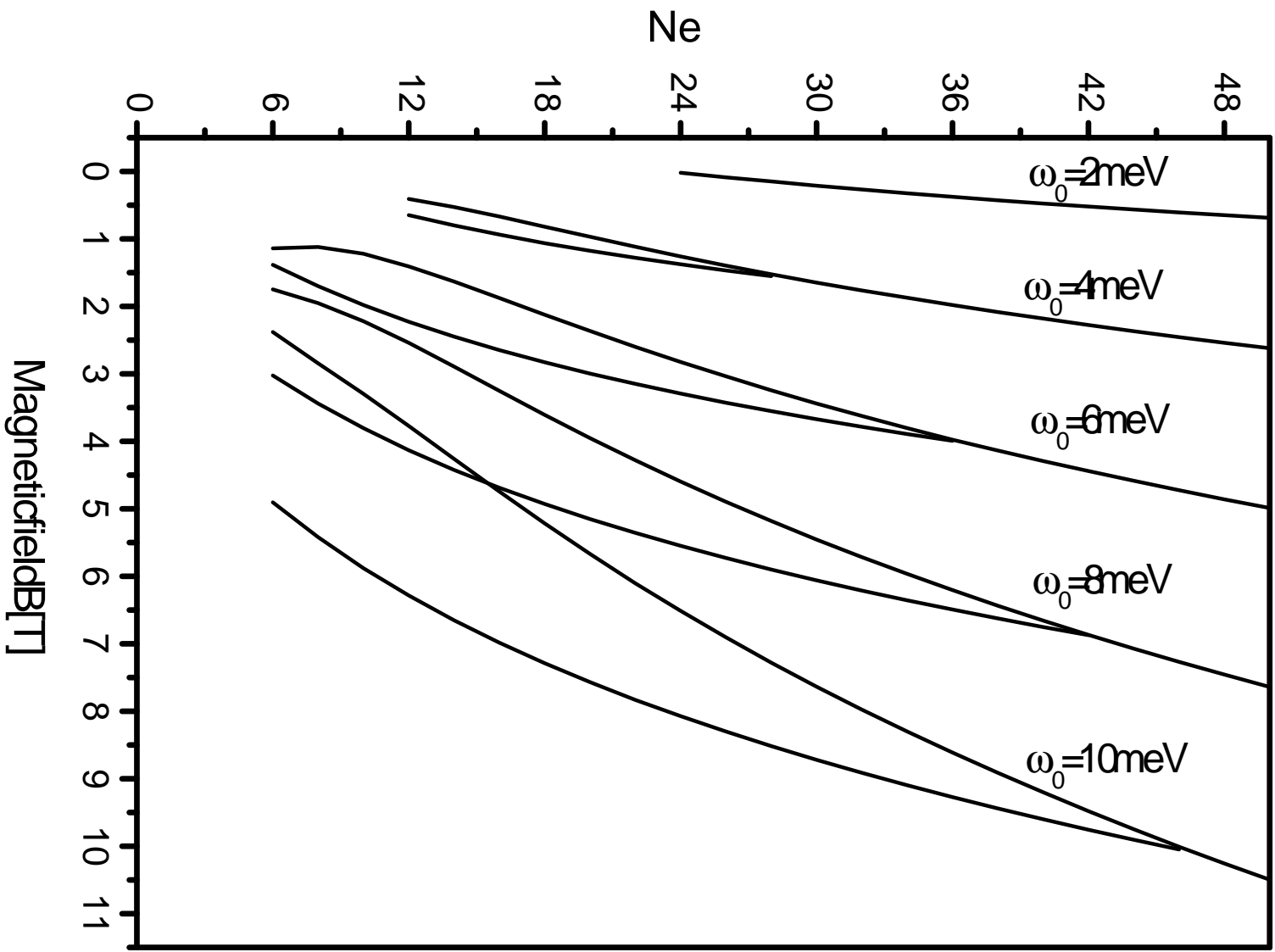
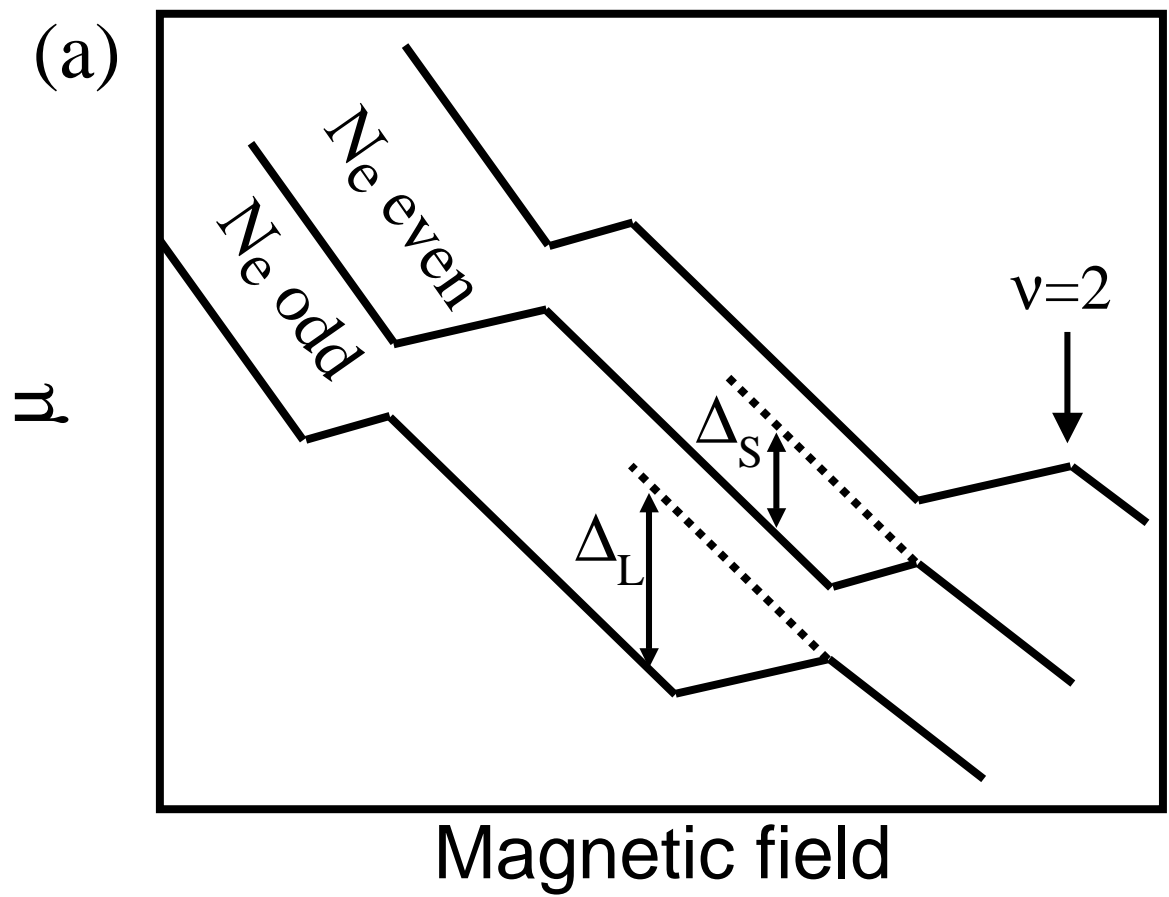
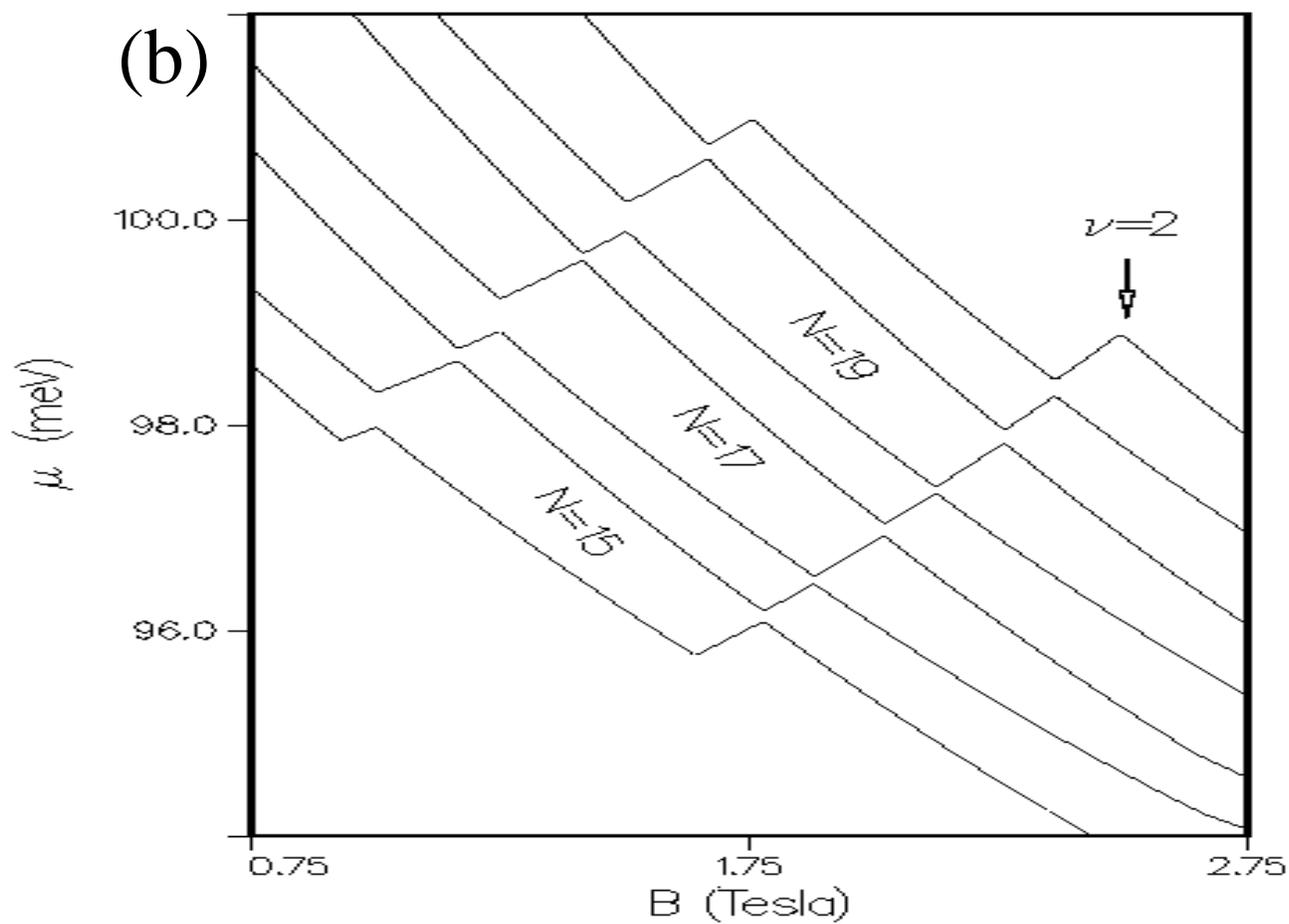
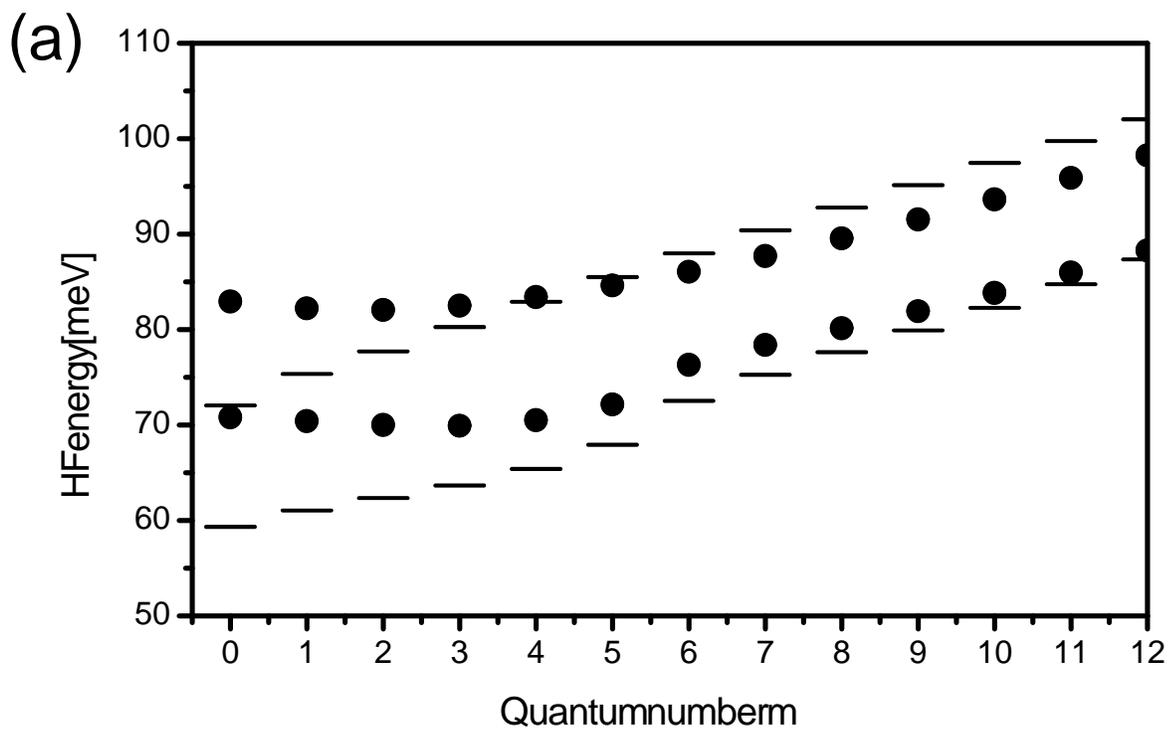
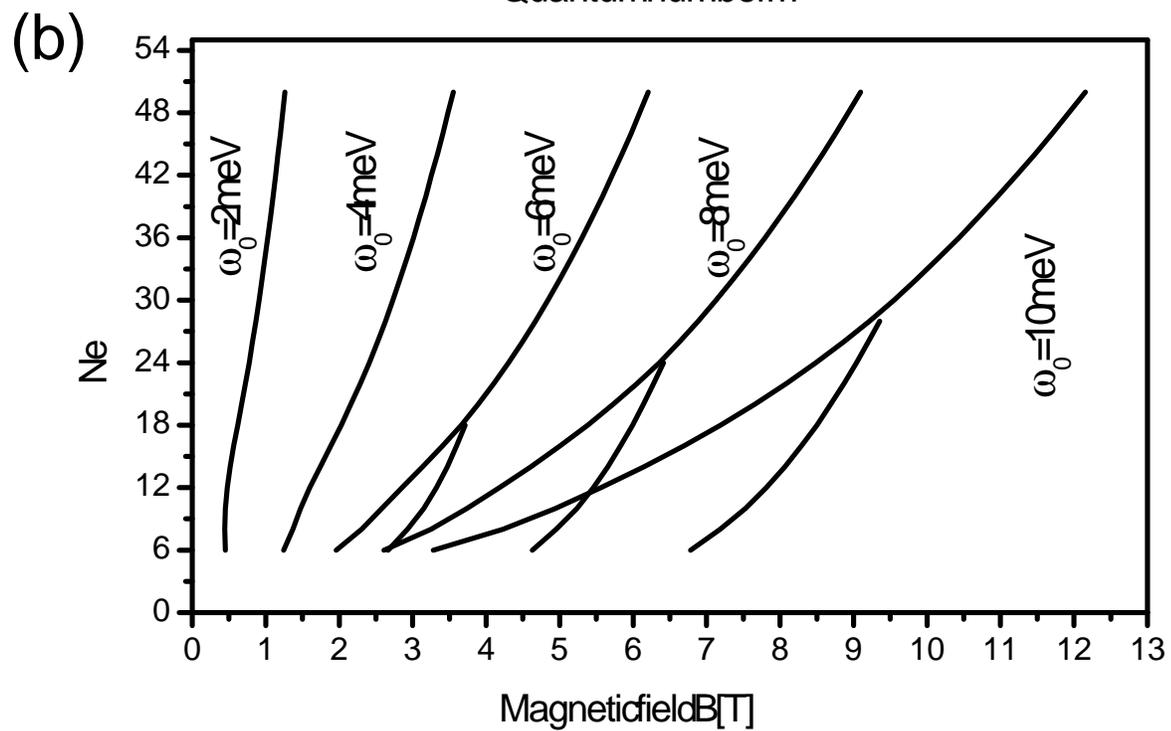
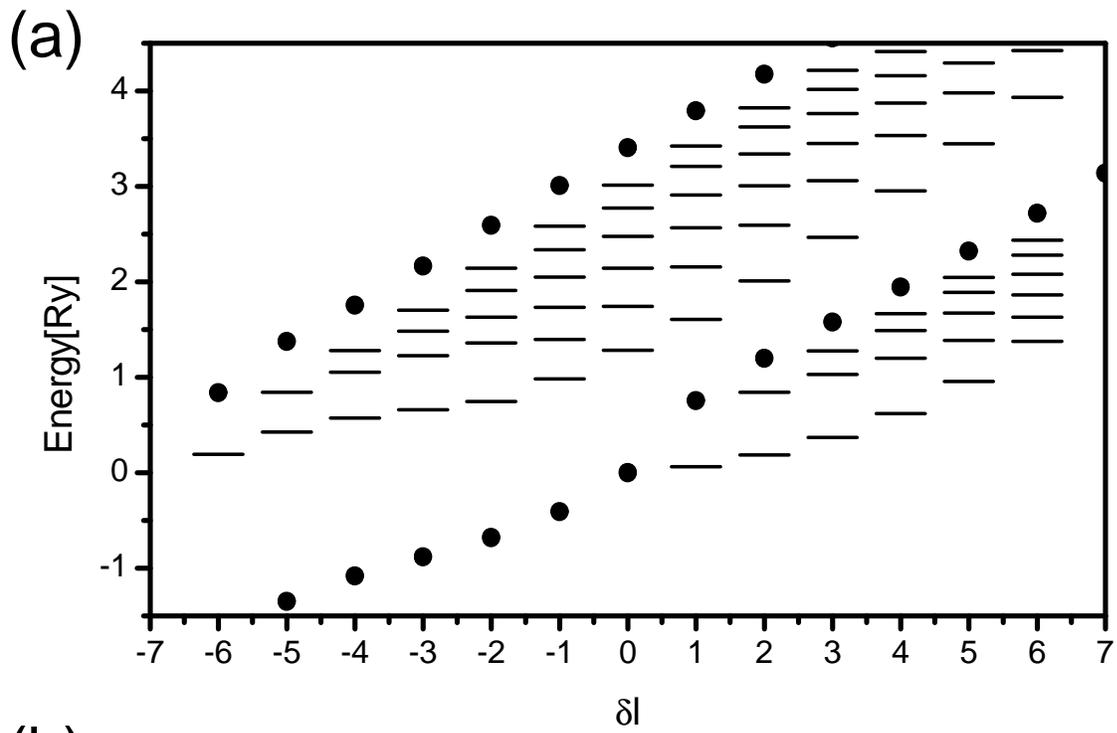
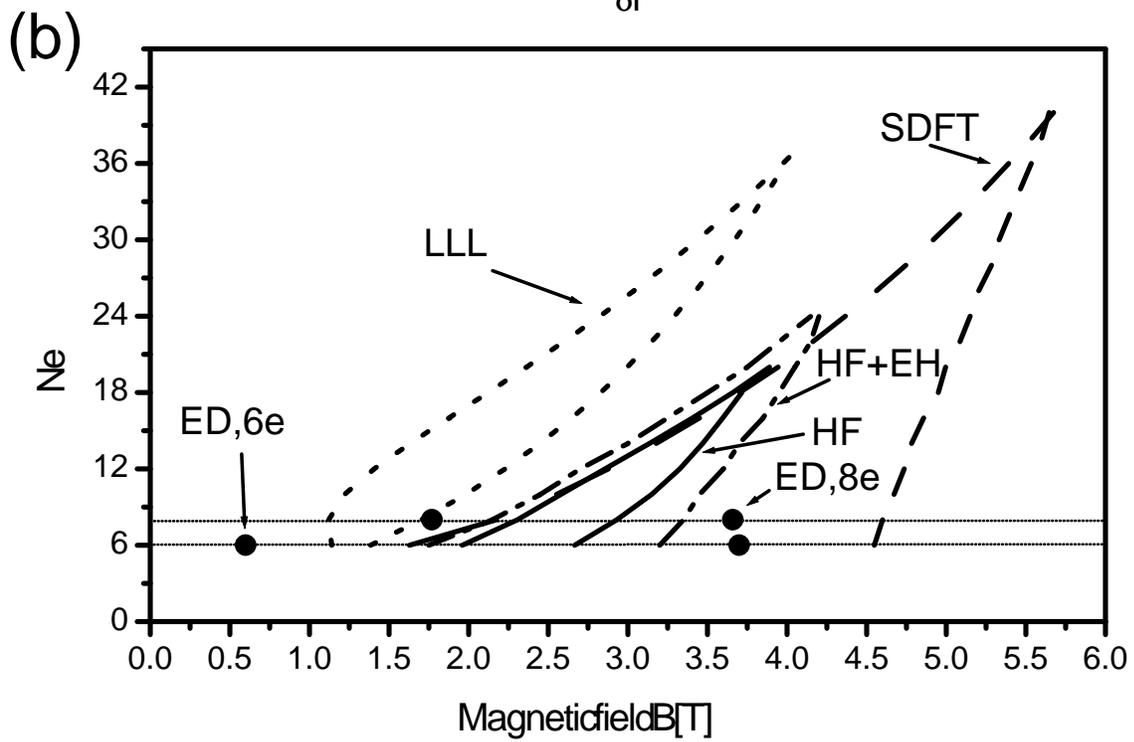
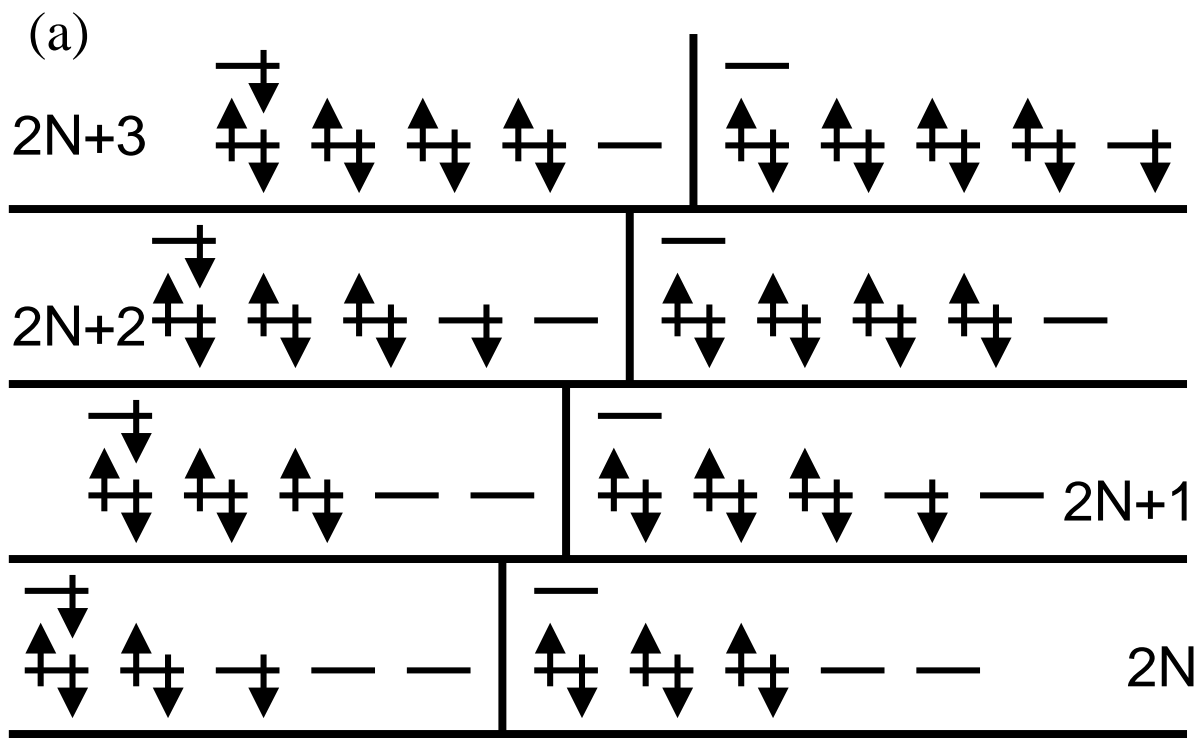
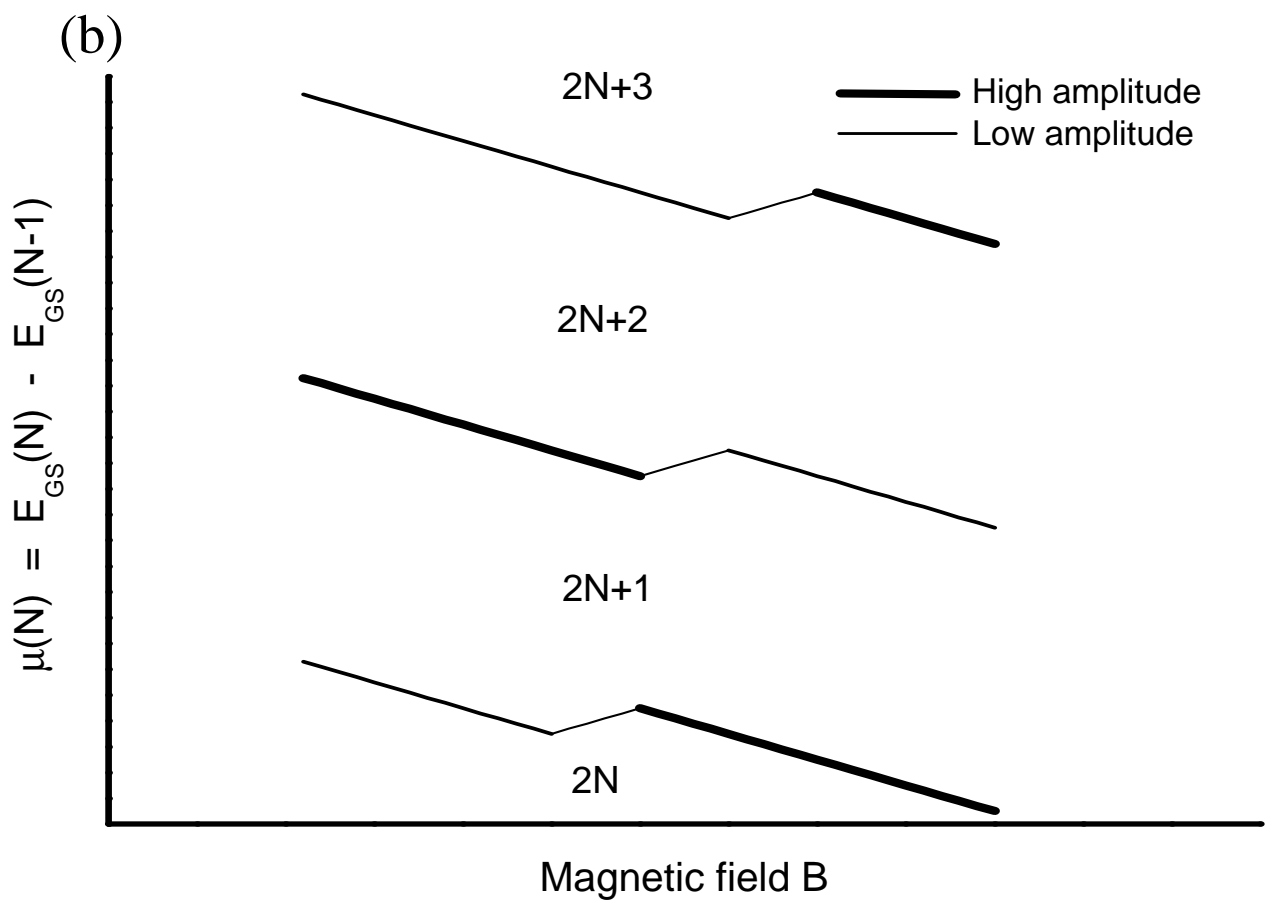
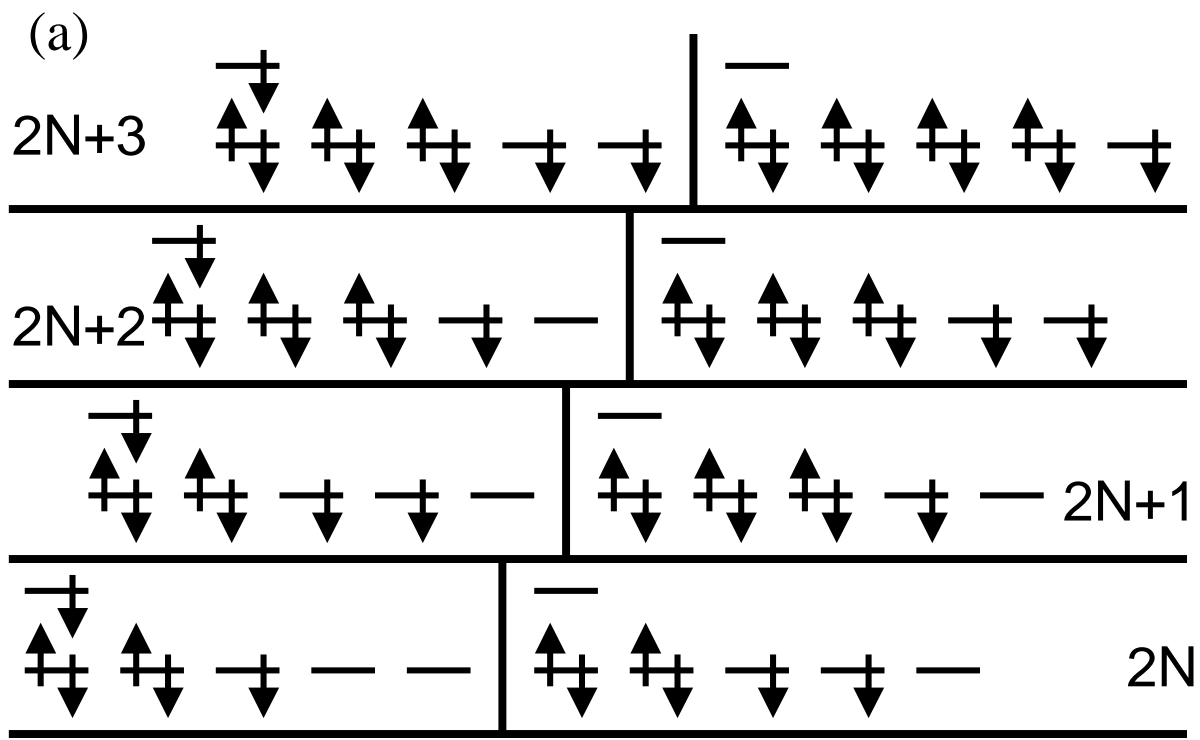
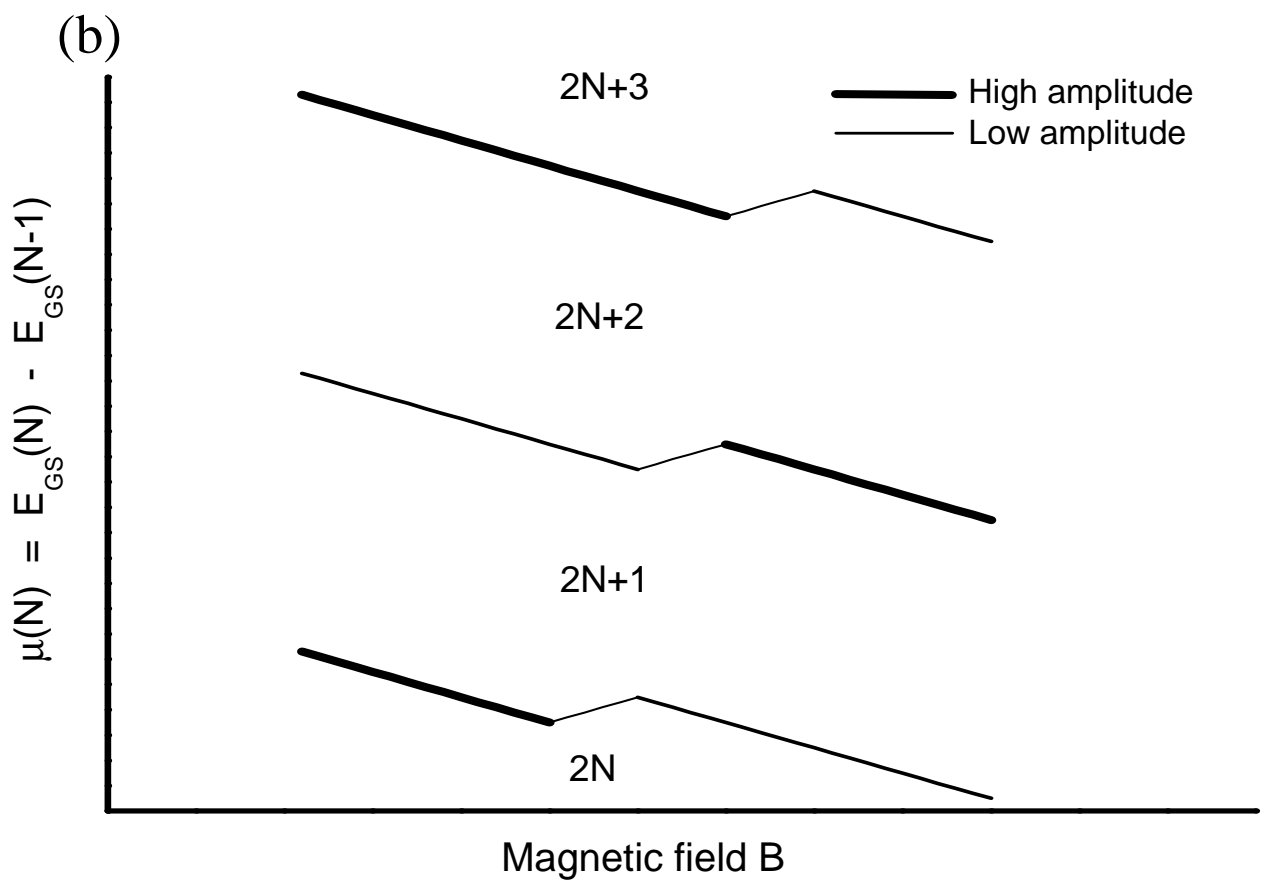
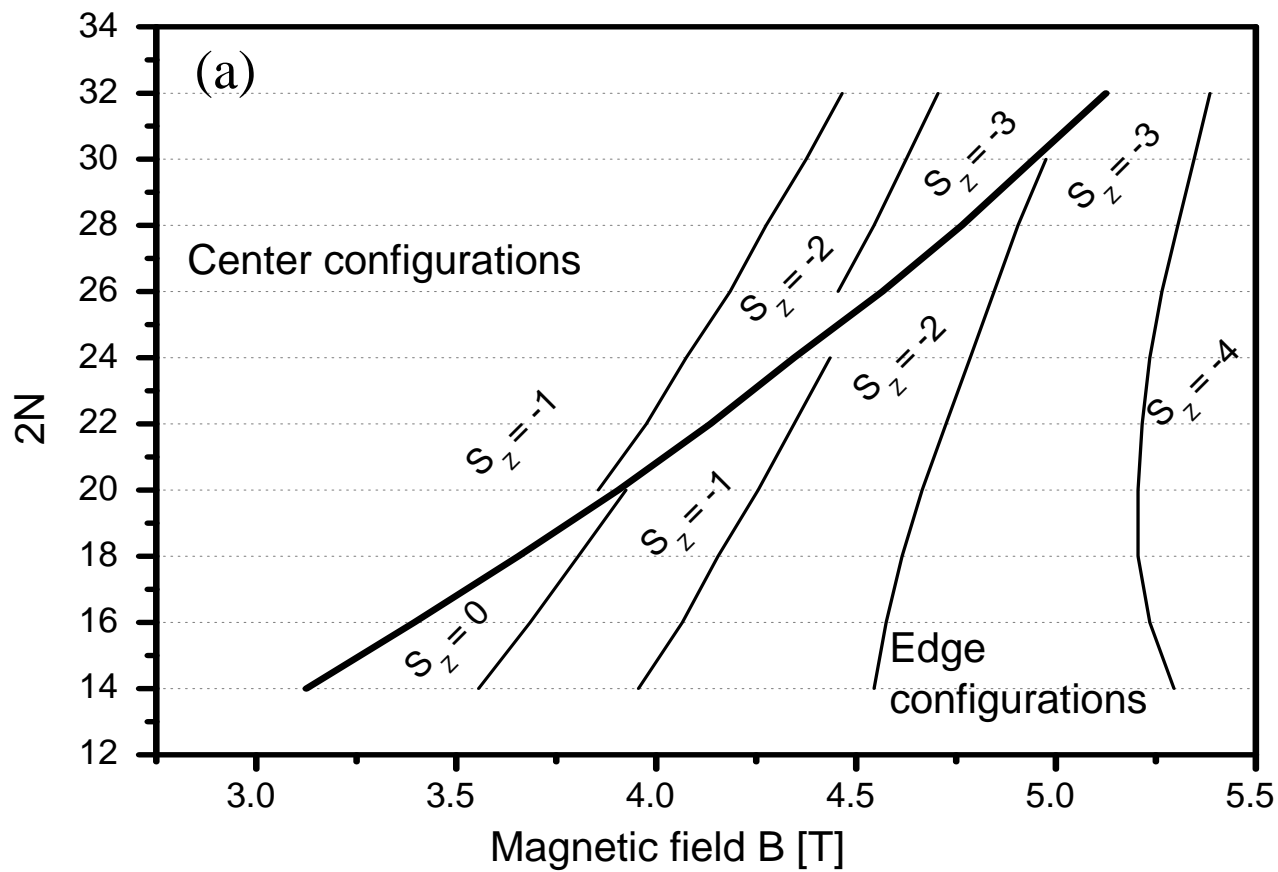
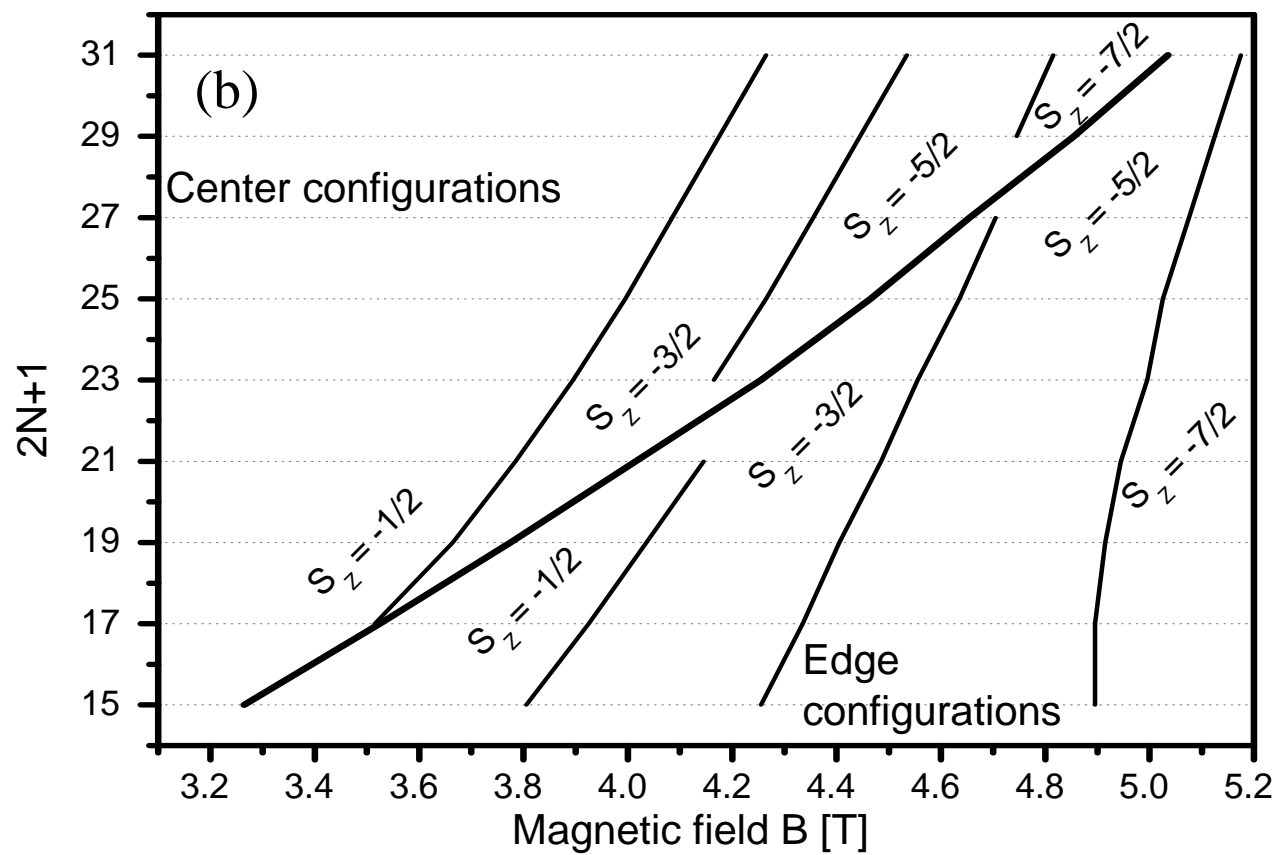
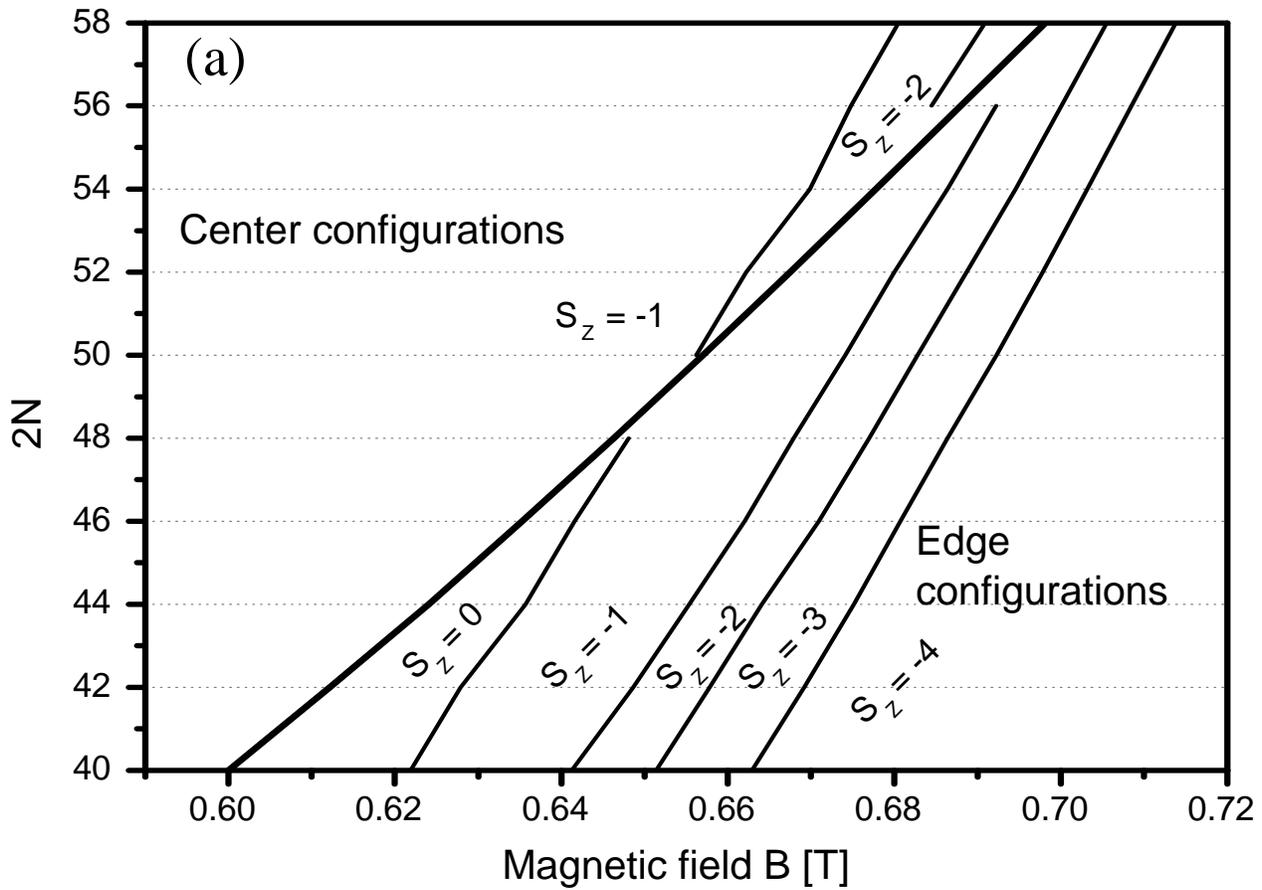
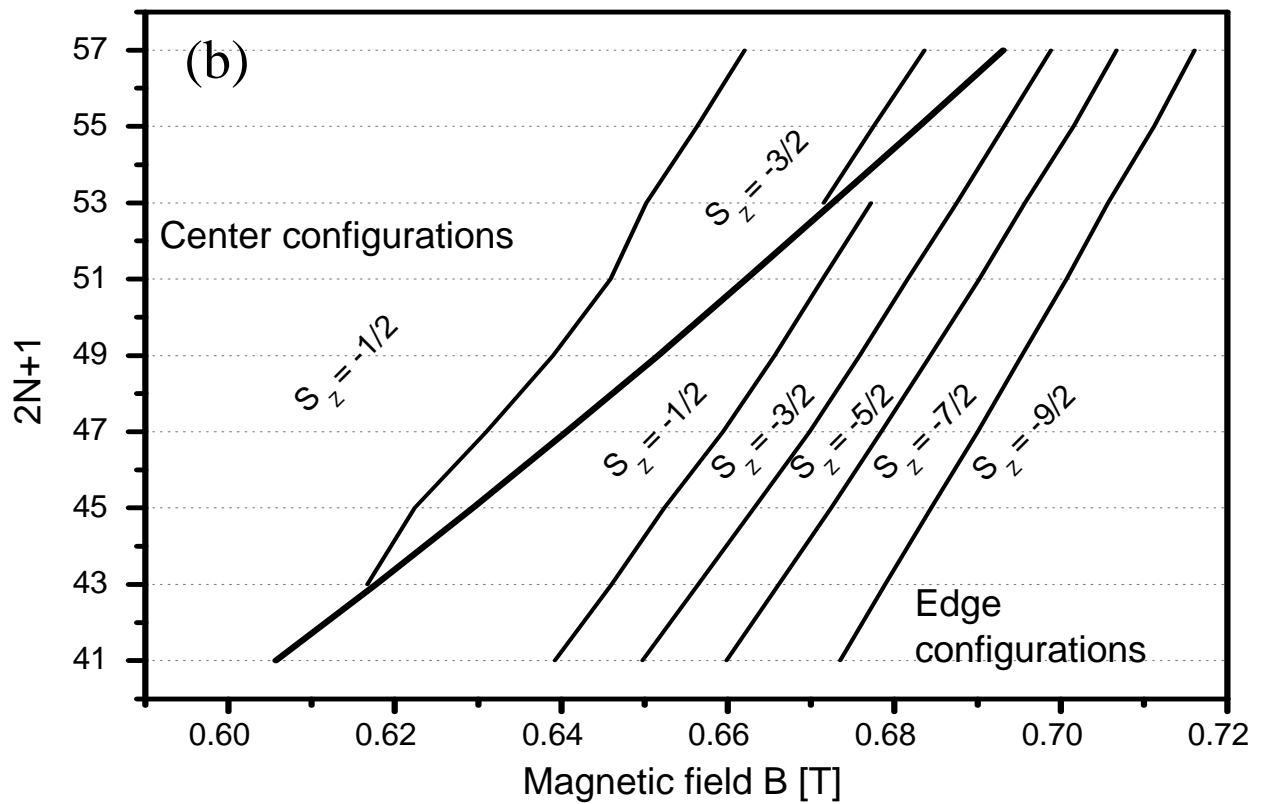
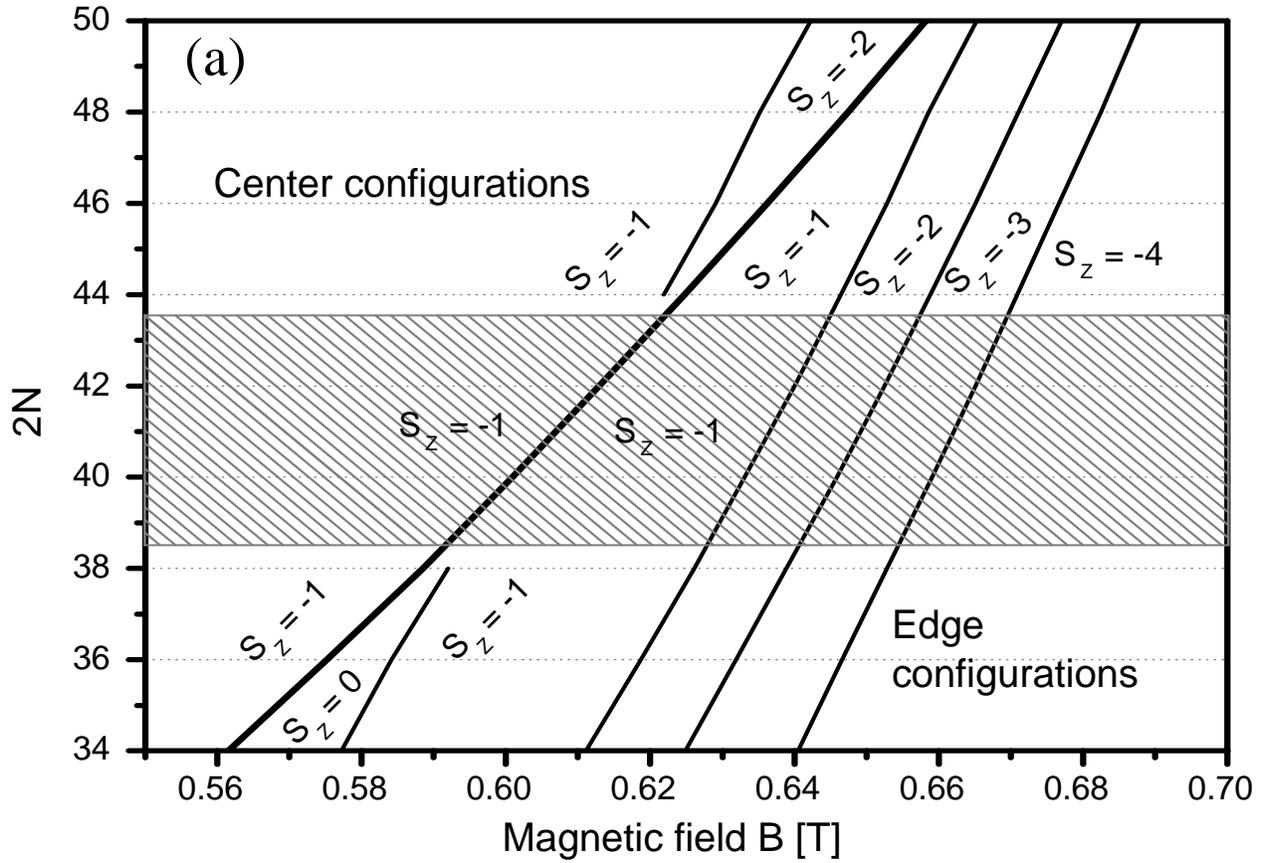
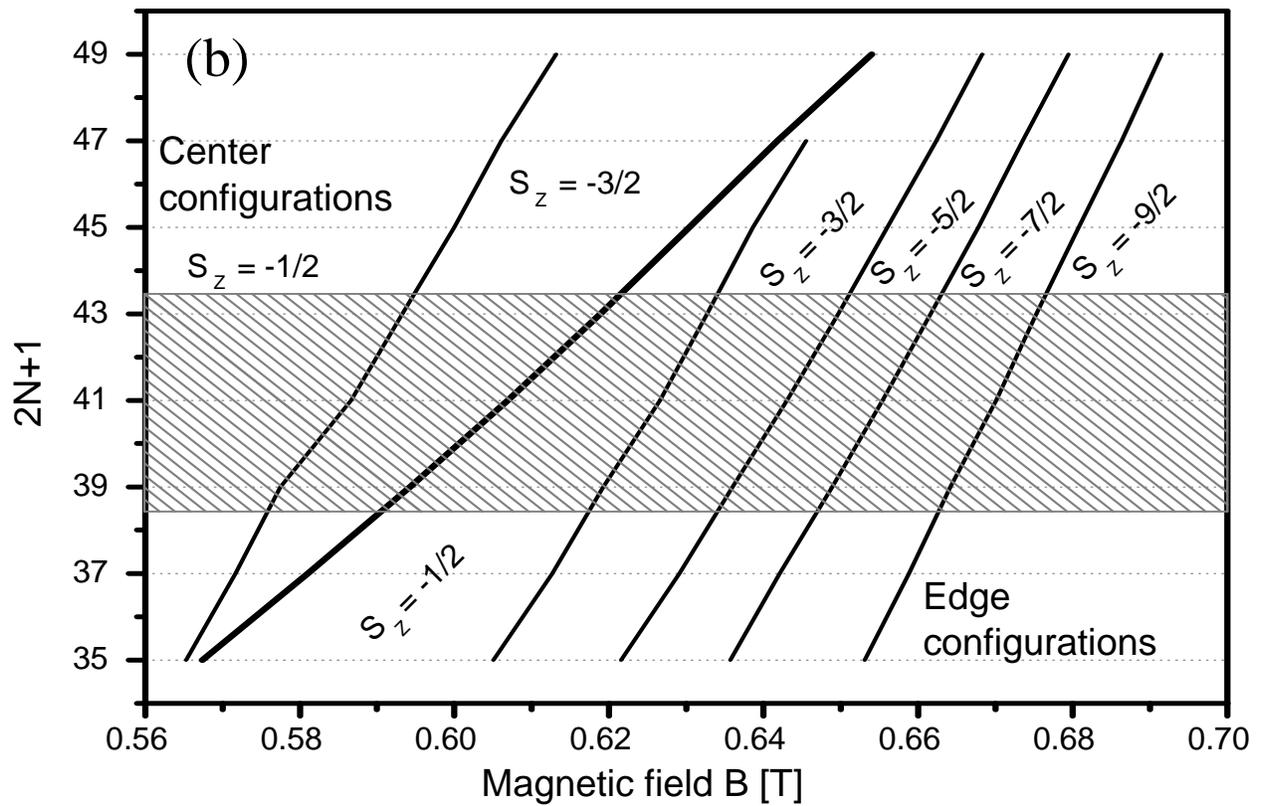
\end{document}